\documentclass[fleqn,usenatbib]{mnras}

\usepackage{newtxtext,newtxmath}

\usepackage[T1]{fontenc}

\DeclareRobustCommand{\VAN}[3]{#2}
\let\VANthebibliography\thebibliography
\def\thebibliography{\DeclareRobustCommand{\VAN}[3]{##3}\VANthebibliography}

\usepackage{graphicx}	
\usepackage{amsmath}	
\usepackage{multirow}
\newlength{\punt}
\newlength{\zero}
\usepackage[dvipsnames]{xcolor}

\title[Physical and numerical viscosity in the KHI]{The role of physical and numerical viscosity in hydrodynamical instabilities}

\author[T. Marin-Gilabert et al.]{
Tirso Marin-Gilabert$^{1}$\thanks{E-mail: tmarin@usm.lmu.de},
Milena Valentini$^{1,2,3}$,
Ulrich P. Steinwandel$^{4}$, 
and Klaus Dolag$^{1,5}$
\\
$^{1}$ Universit{\"a}ts-Sternwarte, Fakult{\"a}t f{\"u}r Physik,  Ludwig-Maximilians-Universit{\"a}t  M{\"u}nchen, Scheinerstr. 1, 81679 M{\"u}nchen, Germany\\
$^{2}$ Excellence Cluster ORIGINS, Boltzmannstr. 2, D-85748 Garching, Germany\\
$^{3}$ INAF - Osservatorio Astronomico di Trieste, via Tiepolo 11, I-34131 Trieste, Italy\\
$^{4}$Center for Computational Astrophysics, Flatiron Institute, 162 5th Avenue, New York, NY 10010\\
$^{5}$Max-Planck-Institut f{\"u}r Astrophysik, Karl-Schwarzschild-Str. 1, D-85741 Garching, Germany
}

\date{Accepted XXX. Received YYY; in original form ZZZ}

\pubyear{2022}

\begin{document}
\label{firstpage}
\pagerange{\pageref{firstpage}--\pageref{lastpage}}
\maketitle

\begin{abstract}
The evolution of the Kelvin-Helmholtz Instability (KHI) is widely used to assess the performance of numerical methods. We employ this instability to test both the smoothed particle hydrodynamics (SPH) and the meshless finite mass (MFM) implementation in \textsc{OpenGadget3}. We quantify the accuracy of SPH and MFM in reproducing the linear growth of the KHI with different numerical and physical set-ups. Among them, we consider: $i)$ numerical induced viscosity, and $ii)$ physically motivated, Braginskii viscosity, and compare their effect on the growth of the KHI. We find that the changes of the inferred numerical viscosity when varying nuisance parameters such as the set-up or the number of neighbours in our SPH code are comparable to the differences obtained when using different hydrodynamical solvers, i.e. MFM. SPH reproduces the expected reduction of the growth rate in the presence of physical viscosity and recovers well the threshold level of physical viscosity needed to fully suppress the instability. In the case of galaxy clusters with a virial temperature of $3\times10^7$~K, this level corresponds to a suppression factor of $\approx10^{-3}$ of the classical Braginskii value. The intrinsic, numerical viscosity of our SPH implementation in such an environment is inferred to be at least an order of magnitude smaller (i.e. $\approx10^ {-4}$), re-ensuring that modern SPH methods are suitable to study the effect of physical viscosity in galaxy clusters.

\end{abstract}

\begin{keywords}
methods: numerical -- hydrodynamics -- turbulence -- instabilities -- galaxies: clusters: intracluster medium 
\end{keywords}



\section{Introduction}

Several astrophysical systems feature a continuous fluid with a velocity shear or two fluids in contact that stream in opposite directions. As time passes, these fluids evolve into a turbulent regime where they mix. This mixing process is primarily driven by the Kelvin-Helmholtz Instability (KHI), where a small perturbation on the interface between the two fluids evolves to an instability with a final curled-up state that leads to a mixture of both fluids. 

The creation of this vortex, a characteristic of the KHI, plays a fundamental role in many environments. Among them, the Intracluster Medium \citep[ICM;][]{Nulsen_1982, Nulsen_1986}, where thermal conduction and turbulence are key processes. The KHI has been observed in cold fronts moving through the ICM \citep[e.g.][]{Breuer_2020, Ge_2020}: it can partially disrupt them, leading to the creation of `bays' on their surface \citep[e.g.][]{Walker_2017}. KHIs can also affect the metallicity profile of galaxy clusters, due to the mixing between the Interstellar Medium (ISM) of galaxies infalling into the cluster and the ICM \citep[e.g.][]{Rebusco_2005}. As a consequence, the ICM gets enriched with metals (from the more metal-rich ISM) and cools outside-in, providing additional gas to form new stars \citep{Muller_2021}.

Although the growth of the KHI can potentially lead to the disruption of a cold front \citep[e.g.][]{Zuhone_2010, Roediger_2011}, several cold fronts are observed to be stable: this suggests that viscosity and magnetic fields can suppress the instability growth \citep[e.g.][]{Vikhlinin_2001, Markevitch_2007}. The viscosity of the ICM plays a fundamental role in the growth of the KHI. Observations of these instabilities provide us with an estimate of the ICM viscosity, even if its actual value is still under debate. By measuring the observed amplitude of the KHI and comparing results with predictions from simulations, the level of suppression experienced by the evolving perturbation can be quantified and hence the ICM viscosity can be estimated. Interestingly, \citet[][]{Roediger_2013_CF} and \citet[][]{Zuhone_2015} analysed the effect that anisotropic viscosity along with magnetic fields has on the suppression of the KHI in cold fronts, without reaching a conclusive agreement. Other works investigated the impact that the ICM viscosity has on the evolution of buoyant bubbles inflated by AGN jets \citep{Sijacki_2006, Dong_2009}. They found that the expanding bubbles might be an important source of heating of the ICM \citep[see also][]{Fabian_2005}. To this end, viscosity must suppress the KHIs so the cavities might remain stable for longer than a crossing time \citep[e.g.][]{Reynolds_2005}. 

Due to its importance in nature, the KHI is commonly used to test numerical codes, such as \textsc{arepo} \citep{Springel_2010}, \textsc{gizmo} \citep{Hopkins_2015} or \textsc{tenet} \cite{Schaal_2015}. In astrophysical simulations, a code that succeeds at reproducing the growth of the KHI is expected to properly capture fluid mixing and turbulence, which is essential when dealing with the evolution of simulated systems.

Different numerical methods have been introduced to perform astrophysical hydrodynamical simulations, which can be mainly divided into two types: Eulerian mesh-based methods \citep[e.g.][]{Evans_1988, Stone_1992}, with and without adaptive mesh refinement \citep[AMR;][]{Berger_1989} and Lagrangian smoothed particle hydrodynamics \citep[SPH;][]{Lucy_1977, Monaghan_1977}. 
In grid-based methods, the volume of the whole domain is discretised into cells, and a Riemann problem is solved between the two states that meet at each plane separating adjacent zones. Solving a Riemann problem produces implicitly entropy when fluxes from different thermodynamic states mix in one single cell: as a consequence, there is no need to add numerical dissipation artificially. However, some finite volume schemes introduce artificial viscosity to stabilize the solver, which is the case of the early version of \textsc{zeus} \citep[][]{Stone_1992}. Nevertheless, mesh-based methods are not strictly Galilean-invariant: therefore, difficulties can arise when e.g. simulating galaxies in high-velocity orbits, where galaxy velocity is larger than the sound speed of the medium \citep[][]{Springel_2010}. In addition, dissipation terms are purely numerical and sensitive to the absolute velocity of the flow, which means that the mixing occurs even when no physical motivation exists \citep[][]{Wadsley_2008}. 

Conversely, in SPH the fluid is discretised in comoving mass elements, leading to Galilean-invariance and exact momentum, energy, angular momentum and entropy conservation. Due to the conservation of entropy, no energy dissipation occurs, which is, in turn, a huge advantage over grid codes. However, this becomes a problem in treating discontinuities and mixing processes properly as a consequence of spurious surface tension at the interface between the two fluids \citep{Agertz_2007}. This was solved by \cite{Price_2008} by introducing an artificial conductivity (AC), which removes the surface tension and allows the mix between the fluids. A physically motivated artificial viscosity (AV) term must be also added \citep{Monaghan_1983, Monaghan_1992} to characterize the Reynolds number in a simulated flow \citep{Wadsley_2008}. As a result, shocks can be treated correctly, removing the post-shock oscillations and noise. Nevertheless, there has been some controversy in the past with SPH schemes due to the fact that they tend to suppress the growth of the KHI \citep{Agertz_2007, McNally_2012}. Increasing the number of neighbours could solve this problem, caused by the `E$_0$ error' \citep[][see also Section~\ref{Kernel functions}, for details]{Read_2010} and which scales sub-linearly with resolution. While this solution does not apply to a cubic spline kernel, it can be useful when adopting Wendland kernels \citep[][]{Wendland_1995, Dehnen_2012}. By increasing the number of neighbours, the KHI can be successfully evolved also with SPH codes \citep{Tricco_Price_2013, Hu_2014}.

To combine the advantages of both grid and SPH schemes, \citet{Gaburov_2011} suggested the idea of mixing both methods based on the mathematical formulation by \citet{Vila_1999} and \citet{Lanson_2008}. This leads to a consistent Lagrangian meshless scheme where the artificial dissipation comes up naturally, like in grid methods. The idea was later developed by \citet{Hopkins_2015} with the introduction of a meshless finite mass (MFM) method based on kernel discretization of the volume and on a higher-order gradient estimator. A weighted kernel is employed, which determines how the volume is partitioned at any point of the fluid among the neighbours to keep the mass constant. Then the fluxes between the particles are computed taking the weighted kernel into account and using a Riemann solver between the particles inside that volume. This allows the particles to move with the flow leading to Galilean-invariance; energy, mass and momentum are conserved and there is no need for artificial diffusion terms \citep[][]{Dave_2016}.

Despite \citet{Moore_1979} proposed an approximation to the exact evolution equation for incompressible fluids, no analytic solution for the non-linear KHI has been achieved yet in the case of compressible fluids. Without an analytical solution, one must rely on reference simulations as an approximation to the true solution. To this end, \citet{Robertson_2010} and \citet{McNally_2012} studied the early linear evolution of the KHI, which was later expanded to the non-linear regime by \citet{Lecoanet_2015} and \citet{Tricco_2019}, setting a benchmark for later comparisons. In their papers, they used smoothed initial conditions to avoid the growth of undesired modes excited by the discontinuity \citep[e.g.][]{Abel_2011, Kawata_2012, Obergaulinger_2020}. In grid codes, the smooth initial conditions are used to suppress truncation errors that can act as seeds of secondary instabilities. This is a particular effect of the second order nature of most Finite Volume reconstruction schemes and typically vanishes when using a lower order reconstruction with larger diffusivity. The cause of this is that, with increasing convergence order in grid codes, edges become sharper and truncation errors become more significant. However, it has been shown that an optimal growth without the seed of secondary instabilities of the KHI can be also achieved by employing discontinuous initial conditions \citep[e.g.][]{Hopkins_2015, Wadsley_2017}. Differently than in grid codes, a higher order construction in SPH typically implies a larger volume that is smoothed, making edges less sharp and smoothing out the truncation error, leading to a natural suppression of secondary instabilities.

The aim of this paper is to study the evolution of the KHI and the mixing processes carried out by this instability during the linear regime. For this purpose, we employ the SPH and the MFM schemes implemented in \textsc{OpenGadget3} and set discontinuous initial conditions to trigger the KHI. We aim at studying how fluid mixing works depending on the code employed in an idealised set-up which allows us to reach a higher resolution than in state-of-the-art cosmological simulations. Additionally, we want to analyse the effect of physical viscosity in these type of processes and study how this could affect the fluid mixing depending on the amount of viscosity implemented.

This paper is organised as follows: In Section~\ref{Theory} we present the equations of hydrodynamics for the ideal and non-ideal cases. In Section~\ref{Numerical_implementation} we describe in detail the numerical methods used in this paper together with the different sets of simulations employed. The results obtained with SPH are shown in Section~\ref{SPH_results}. In Section~\ref{MFM_results} we compare the results obtained with SPH with the ones obtained using MFM. Once we have deeply analysed the results obtained for ideal fluids, in Section~\ref{Physical_viscosity} we study how the addition of physical viscosity affects the previous results. Finally, we test different initial conditions in Section~\ref{Read_IC} to see the effect they have in triggering the KHI.

\section{Theoretical Considerations} \label{Theory}

\subsection{Equations of fluids}

Equations of hydrodynamics rule the motion of fluids. These equations describe the conservation of mass, momentum and energy. In the case of inviscid fluids, they can be written as:
\begin{equation}
    \frac{\partial \rho}{\partial t} + \frac{\partial (\rho v_k)}{\partial x_k} = 0 \, ,
    \label{eqn:continuity_eq_tensor}
\end{equation}
\begin{equation}
    \frac{\partial (\rho v_i)}{\partial t} + \frac{\partial}{\partial x_k}\left(\rho v_i v_k + \delta_{ik} P\right) = 0 \, ,
    \label{eqn:momentum_eq_tensor}
\end{equation}
\begin{equation}
     \frac{\partial (\rho e)}{\partial t} + \frac{\partial}{\partial x_k} \left[ \left(\rho e + P \right) \, v_k \right] = 0 \, .
     \label{eqn:energy_eq_tensor}
\end{equation}
Here, $\rho$ is the fluid density, $v$ the velocity, $P$ the pressure, $e = u + \frac{1}{2} v^2$ the total energy per unit mass and $u$ the specific internal energy. For an ideal gas, the pressure is related to the density via the equation of state 
\begin{equation}
    P = (\gamma - 1) \, \rho \, u \, ,
    \label{eqn:equation_state}
\end{equation}
$\gamma$ being the adiabatic index. In the case of an ideal monoatomic gas $\gamma = 5/3$.

However, real fluids are viscous and the amount of their viscosity determines their properties and behaviour. Hydrodynamic equations of ideal fluids must be modified to account for the friction between particles of viscous fluids. The continuity equation \ref{eqn:continuity_eq_tensor} does not change, whereas the momentum and the heat transfer equations are altered \citep[see e.g.][]{Landau_1987}.

The viscosity-induced friction between particles leads to a change of their momentum. This is given by an additional term in equation \ref{eqn:momentum_eq_tensor}, which becomes:
\begin{equation}
    \frac{\partial (\rho v_i)}{\partial t} + \frac{\partial}{\partial x_k}\left(\rho v_i v_k + \delta_{ik} P\right) = \frac{\partial \sigma_{ik}}{\partial x_k} \, .
    \label{eqn:momentum_eq_visc}
\end{equation}
Here, $\sigma_{ik}$ is the viscous stress tensor defined as
\begin{equation}
    \sigma_{ik} = \eta \left( \frac{\partial v_{i}}{\partial x_{k}} + \frac{\partial v_{k}}{\partial x_{i}} - \frac{2}{3} \delta_{ik} \frac{\partial v_{l}}{\partial x_{l}} \right) + \zeta \delta_{ik} \frac{\partial v_{l}}{\partial x_{l}} \, .
    \label{eqn:sigma}
\end{equation}
The second term on the right-hand side is the bulk viscosity term, where $\zeta$ is the bulk viscosity coefficient. Since it only depends on the divergence of the velocity, it becomes relevant when there is a rapid compression or expansion of the fluid, i.e. shocks. The first term on the right-hand side is the shear viscosity term and $\eta$ is the shear viscosity coefficient (dynamic viscosity). This was derived by Braginskii \citep{Braginskii_1958, Braginskii_1965} for a fully ionized, unmagnetized plasma and reads:
\begin{equation}
    \eta = 0.406 \frac{m_i^{1/2} T_i^{5/2}}{(Ze)^4 \ln \Lambda} \, ,
    \label{eqn:shear_visc_coeff}
\end{equation}
where $m_i$ is the mass of the proton, $T_i$ is the temperature of the plasma, $Ze$ is the ion charge and $\ln \Lambda$ is the Coulomb logarithm. 
Plugging equation \ref{eqn:sigma} into \ref{eqn:momentum_eq_visc} we get:
\begin{equation}
    \rho \left( \frac{\partial v_i}{\partial t} + v_k \frac{\partial v_i}{\partial x_k}\right) = - \frac{\partial P}{\partial x_i} + \frac{\partial \sigma_{ik}}{\partial x_k} \, ,
    \label{eqn:momentum_eq_visc2}
\end{equation}
\begin{multline}
    \rho \left( \frac{\partial v_i}{\partial t} + v_k \frac{\partial v_i}{\partial x_k}\right) = - \frac{\partial P}{\partial x_i} + \frac{\partial}{\partial x_k} \left[ \eta \left( \frac{\partial v_{i}}{\partial x_{k}} + \frac{\partial v_{k}}{\partial x_{i}} - \frac{2}{3} \delta_{ik} \frac{\partial v_{l}}{\partial x_{l}} \right) \right] + \\
    + \frac{\partial}{\partial x_i} \left( \zeta \frac{\partial v_l}{\partial x_l} \right) \, .
\end{multline}
Both $\eta$ and $\zeta$ are positive functions of $P$ and of the temperature $T$: since they vary along the fluid, $\eta$ and $\zeta$ cannot be taken off the partial derivatives.

Equation \ref{eqn:energy_eq_tensor} can be written in vector form as
\begin{equation}
    \frac{\partial (\rho e)}{\partial t} = -\nabla \cdot \left[ \left( \rho e + P \right) \, \mathbf{v} \right] \, .
\end{equation}
The internal friction also contributes to the energy flux density (right-hand side). This contribution is $\mathbf{v} \cdot \mathbf{\sigma}$. If the temperature is not constant, there is also a heat transfer due to thermal conduction, which reads $\kappa \nabla T$, where $\kappa$ is the coefficient of thermal conductivity \citep[e.g.][]{Landau_1987}. By adding those terms, equation \ref{eqn:energy_eq_tensor} becomes: 
\begin{equation}
    \frac{\partial (\rho e)}{\partial t} = - \nabla \cdot \left[ \left( \rho e + P \right) \, \mathbf{v} - \mathbf{v} \cdot \mathbf{\sigma} - \kappa \nabla T \right] \, .
\end{equation}

\subsection{Kelvin-Helmholtz Instability}\label{KHI}

A 2D linear analysis of the KHI \citep[e.g.][]{Junk_2010} shows that the $y$-velocity of the perturbation grows exponentially $\sim \exp{[i \cdot n \cdot t]}$. $n$ is the mode of the perturbation:
\begin{equation}
    n = \left[ k^2 v_x^2 (\alpha_2 - \alpha_1) \right] + i \left[\frac{\nu k^2}{2} \pm \sqrt{\frac{\nu^2 k^4}{4} + 4k^2 v_x^2 \alpha_1 \alpha_2} \right] \, ,
    \label{eqn:mode_perturbation}
\end{equation}
where $k$ is the wavenumber of the perturbation, $v_x$ is the velocity of one of the fluids (in the laboratory frame of reference) and $\nu$ is the kinematic viscosity. $\alpha_1$ and $\alpha_2$ are defined as
\begin{equation}
    \alpha_1 = \frac{\rho_1}{\rho_1 + \rho_2} \, , \hspace{2cm} \alpha_2 = \frac{\rho_2}{\rho_1 + \rho_2} \, .
\end{equation}
The real part of equation \ref{eqn:mode_perturbation} deals with the oscillatory behaviour of the KHI and is not of interest here. The imaginary part determines whether the KHI decays (positive solution of the square root) or grows exponentially (negative solution), damped by $\nu$.

In the ideal case where $\nu = 0$, equation \ref{eqn:mode_perturbation} becomes:
\begin{equation}
    n = \left[ k^2 v_x^2 (\alpha_2 - \alpha_1) \right] + i \left[\pm2k v_x (\alpha_1 \alpha_2)^{1/2} \right] \, .
    \label{eqn:mode_perturbation_ideal}
\end{equation}
Expressing the exponential growth of the perturbation as $\sim \exp{[i \omega t]}$, one can write the growth time of the KHI as
\begin{equation}
    \tau_\mathrm{KH} = \frac{2\pi}{\omega} = \frac{2\pi}{2k v_x (\alpha_1 \alpha_2)^{1/2}} = \frac{\lambda}{\Delta v_x} \frac{(\rho_1 + \rho_2)}{(\rho_1\,\rho_2)^{1/2}} \, ,
    \label{eqn:KH_growth_time}
\end{equation}
where $\Delta v_x$ is the velocity difference between the two fluids ($\Delta v_x = 2 v_x$) and $\lambda = \frac{2\pi}{k}$ the wavelength of the perturbation.

\section{Numerical Implementation} \label{Numerical_implementation}

\subsection{Set-up} \label{set-up}

\begin{figure}
	\includegraphics[width=\columnwidth]{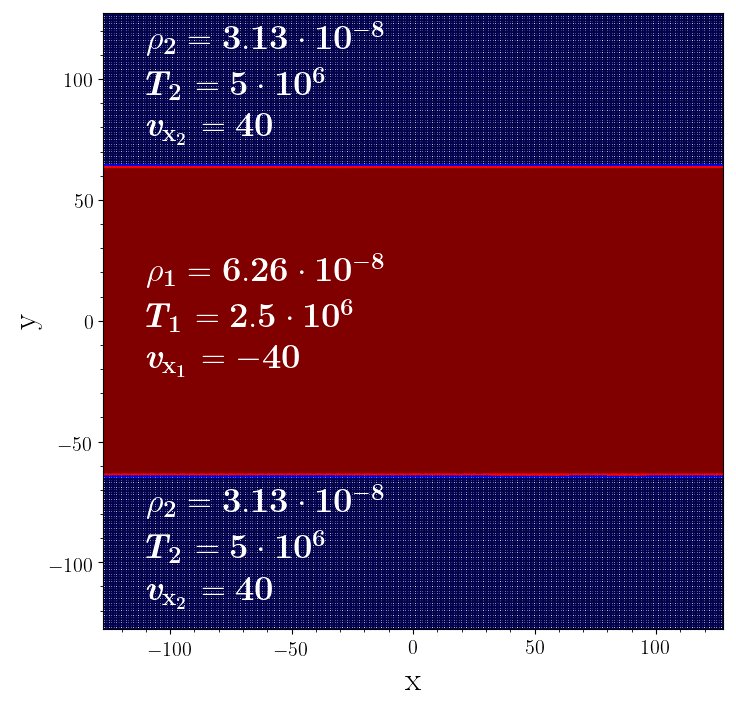}
    \caption{Graphical description of the ICs set in our simulations.}
    \label{fig:ICs_sketch}
\end{figure}

We set up the initial conditions (ICs) following \cite{Murante_2011}. We create a 3D box (see Fig.~\ref{fig:ICs_sketch}) with 774144 particles of equal mass ($m = 3.13\cdot10^{-8}$), using a cubic lattice with periodic boundary conditions. The size of the box in internal units\footnote{The internal units correspond to the basic Gadget units, where mass is given in $10^{10}$~M$_\odot$, length in kpc and velocity in km/s. We use a mean molecular weight of $\approx 0.588$. In this paper we will always refer to internal units.} is $\Delta x = 256$, $\Delta y = 256$, and $\Delta z = 8$. The domain satisfies:
\begin{equation}
    \rho, T, v_x = \left \{ 
    \begin{matrix} \rho_1, T_1, v_1 & |y| < 64 \\
    \rho_2, T_2, v_2 & |y| > 64 \end{matrix} \right .
\end{equation}
Densities, temperatures, and $x$-velocities are: $\rho_1 = 6.26\cdot10^{-8}$, $\rho_2 = 3.13\cdot10^{-8}$; $T_1 = 2.5\cdot10^6$, $T_2 = 5\cdot10^6$; $v_1 = -40$ and $v_2 = 40$, respectively. The density and temperature ratio is constant, i.e. $R_\rho = \rho_1 / \rho_2 = T_2 / T_1 = 2$, ensuring a pressure equilibrium in the system. With these ICs, the Mach number corresponds to $M_1 = v_1/c_1 \approx 0.23$ for the first fluid and $M_2 = -v_2/c_2 \approx 0.17$ for the second one. The initial properties of both fluids can be seen in Fig.~\ref{fig:ICs_sketch}. 

We introduce a small perturbation in the $y$-velocity at $y_{\mathrm{Int}} = \pm 64$ (equation \ref{eqn:vy}) to trigger the instability. This is similar to \cite{Read_2010}, but values have been adapted to our ICs:
\begin{multline}
    v_y = -\delta v_y \left[ \sin \left( \frac{2\pi (x+\lambda/2)}{\lambda}\right) \exp \left(-\left(\frac{y-y_{\mathrm{Int}}}{\sigma}\right)^2\right) + \right. \\
    \left. + \sin \left( \frac{2\pi x}{\lambda}\right) \exp \left(-\left(\frac{y+y_{\mathrm{Int}}}{\sigma}\right)^2\right) \right] \, .
    \label{eqn:vy}
\end{multline}
Here, $\lambda = 128$ is the wavelength of the perturbation, $\delta v_y = |v_x|/10 = 4$ is the amplitude and $\sigma = 0.2\lambda$ is a scaling parameter to control the width of the perturbation layer. Table \ref{tab:Read_IC} summarizes the set-up.

\subsection{Artificial diffusion in SPH}

We have mitigated difficulties of traditional SPH schemes in treating contact discontinuities and shocks (due to the SPH entropy conserving nature), by implementing artificial diffusion mechanisms \citep[e.g.][]{Monaghan_1983, Monaghan_1992, Price_2008}, as detailed below. 

\subsubsection{Artificial conductivity}

The artificial conductivity \citep[AC;][]{Price_2008} is introduced in \textsc{OpenGadget3} to treat discontinuities in the internal energy and capture mixing processes properly \citep[we follow the notation introduced by][]{Beck_2015}. The AC implemented can be time dependent (labeled in our simulations as ``TDAC''), where the variation of internal energy due to AC reads:
\begin{equation}
    \left. \frac{\mathrm{d}u_i}{\mathrm{d}t} \right|_{\mathrm{cond}} = \sum\limits_{j=1}^N \frac{m_j}{\rho_{ij}} (u_j - u_i) \, \alpha_{ij}^c v_{ij}^{\mathrm{sig,}c} \overline{F}_{ij} \, .
\end{equation}
Here, $v_{ij}^{\mathrm{sig,}c}$ is the signal velocity, $\overline{F}_{ij} = (F_{ij}(h_i) + F_{ij}(h_j))/2$ is the symmetrised scalar part of the kernel gradient terms $\nabla_{i} W_{ij}(h_i) = F_{ij} \, \hat{r}_{ij}$, $\rho_{ij} = (\rho_i + \rho_j)/2$ is the symmetrised density and $\alpha_{ij}^c = (\alpha_i^c + \alpha_j^c)/2$ is the symmetrised conduction coefficient. 
The sum spreads over $N=N_{\mathrm{ngb}}$, the number of neighbours.
The signal velocity depends on the pressure gradient \citep[see][]{Price_2008}:
\begin{equation}
    v_{ij}^{\mathrm{sig,}c} = \sqrt{\frac{|P_i - P_j|}{\rho_{ij}}} \, .
    \label{eqn:signal_vel_AC}
\end{equation}
The AC coefficient is defined as 
\begin{equation}
    \alpha_i^c = \frac{h_i}{2} \frac{|\nabla u|_i}{|u_i|} \, ,
\end{equation}
where the time dependence stems from the dependence on the internal energy and on its gradient, which is computed using
\begin{equation}
    (\nabla u)_i = \frac{1}{\rho_i} \sum\limits_{j=1}^N m_j \, (u_j - u_i) \nabla_i W_{ij} \, .
\end{equation}
When $\alpha_i^c$ is larger than a threshold value (in our case $\alpha_{\mathrm{max}} = 1.0$), we set $\alpha_i^c = \alpha_{\mathrm{max}}$.
 
Simulations which feature a constant AC have $\alpha_i^c = 1.0$ (see table~\ref{tab:set_simulations}).

\subsubsection{Artificial viscosity}

\textsc{OpenGadget3} includes artificial viscosity (AV), to damp post-shocks oscillations and reduce kernel distribution noise. Since in ideal fluids AV is not needed away from shocks, a switch triggers viscosity in shocks but keeps it inactive otherwise. Our implementation of an adaptive AV reads:
\begin{equation}
    \left. \frac{\mathrm{d}\mathbf{v_i}}{\mathrm{d}t} \right|_{\mathrm{visc}} = \frac{1}{2} \sum\limits_{j=1}^N \frac{m_j}{\rho_{ij}} \, (\mathbf{v_j} - \mathbf{v_i}) \, \alpha_{ij}^v \, f_{ij}^{\mathrm{shear}} \, v_{ij}^{\mathrm{sig, v}} \overline{F}_{ij} \, .
\end{equation}

An additional term accounting for the variation of internal energy due to AV offsets the work done against the viscous force in the thermal reservoir. It reads:
\begin{equation}
    \left. \frac{\mathrm{d}u_i}{\mathrm{d}t} \right|_{\mathrm{visc}} = - \frac{1}{2} \sum\limits_{j=1}^N \frac{m_j}{\rho_{ij}} \, (\mathbf{v_j} - \mathbf{v_i})^2 \, \alpha_{ij}^v \, f_{ij}^{\mathrm{shear}} \, v_{ij}^{\mathrm{sig, v}} \overline{F}_{ij} \, .
\end{equation}
Here, $\alpha_{ij}^v = (\alpha_i^v + \alpha_j^v)/2$ is the symmetrised viscosity coefficient, $f_{ij}^{\mathrm{shear}} = (f_i^{\mathrm{shear}} + f_j^{\mathrm{shear}})/2$ the symmetrised shear flow limiter and $v_{ij}^{\mathrm{sig, v}}$ the pairwise signal velocity.

The signal velocity\footnote{Note that this signal velocity is not the same as for the AC (\ref{eqn:signal_vel_AC}).} \citep{Monaghan_1997} aids in switching on or off the AV, depending on whether two particles are approaching ($\mathbf{v_{ij}} \cdot \hat{r}_{ij} \leq 0$) or moving away ($\mathbf{v_{ij}} \cdot \hat{r}_{ij} > 0$), respectively. It also determines the strength of the AV and measures the particle disorder:
\begin{equation}
    v_{ij}^{\mathrm{sig, v}} =
    \begin{cases}
        c_{\mathrm{s}, i} + c_{\mathrm{s}, j} - \beta \, \mathbf{v_{ij}} \cdot \hat{r}_{ij} \, , & \mathbf{v_{ij}} \cdot \hat{r}_{ij} \leq 0 ; \\
        0 \, , & \mathbf{v_{ij}} \cdot \hat{r}_{ij} > 0 ,
    \end{cases}
\end{equation}
where $c_{\mathrm{s}}$ is the sound speed of the particle and $\beta =3$ \citep[see][]{Beck_2015}.

To avoid a shear viscosity that could lead to sub-optimum behaviour in simulations of shear flows, \cite{Balsara_1995} suggested the shear flow limiter:
\begin{equation}
    f_i^{\mathrm{shear}} = \frac{| \nabla \cdot \mathbf{v} |_i}{| \nabla \cdot \mathbf{v} |_i + | \nabla \times \mathbf{v} |_i + \sigma_i} \, ,
\end{equation}
with $\sigma_i = 0.0001 c_{\mathrm{s}, i} / h_i$ for numerical stability reasons. When there is a shock, the limiter is dominated by $| \nabla \cdot \mathbf{v} |_i$ and thus, $f_i^{\mathrm{shear}} \simeq 1$, while if there is a shearing flow, the limiter is dominated by $| \nabla \times \mathbf{v} |_i$ and $f_i^{\mathrm{shear}} \simeq 0$. 

The viscosity coefficient $\alpha_i^v$ is computed following \cite{Cullen_2010}, which use the shock indicator
\begin{equation}
    R_i = \frac{1}{\rho_i} \sum\limits_{j=1}^N \mathrm{sign} (\nabla \cdot \mathbf{v})_j m_j W_{ij} \, ,
\end{equation}
where $\mathrm{sign} (\nabla \cdot \mathbf{v})_j$ is negative and, therefore, $R_i \simeq -1$ when there is a shock. Nevertheless, $R_i$ cannot distinguish between pre- and post-shock regions. To determine the direction of the shock, an additional factor, $A_i$, exploits the time derivative of the velocity divergence:
\begin{equation}
    A_i = \xi_i \, \mathrm{max}(0, -(\dot{\nabla} \cdot \mathbf{v})_i) \, ,
\end{equation}
where $(\dot{\nabla} \cdot \mathbf{v})_i < 0$ indicates a pre-shock region and $(\dot{\nabla} \cdot \mathbf{v})_i > 0$ a post-shock region. $\xi_i$ indicates the ratio of strength of the shock and reads:
\begin{equation}
    \xi_i = \frac{\left| 2(1-R_i)^4 (\nabla \cdot \mathbf{v})_i \right|^2}{\left| 2(1-R_i)^4 (\nabla \cdot \mathbf{v})_i \right|^2 + \left| \nabla \times \mathbf{v}\right|_i^2} \, .
\end{equation}
The target value $\alpha_i^{\mathrm{loc,}v}$ of AV is therefore:
\begin{equation}
    \alpha_i^{\mathrm{loc,}v} = \alpha_{\mathrm{max}} \frac{h_i^2 A_i}{h_i^2 A_i + (v_i^{\mathrm{sig}})^2} \, .
\end{equation}
When $\alpha_i^{\mathrm{loc,}v} < \alpha_i^v$, the latter is set to $\alpha_i^{\mathrm{loc,}v}$. If $\alpha_i^{\mathrm{loc,}v} > \alpha_i^v$, $\alpha_i^v$ decays with time and the value is calculated by integrating:
\begin{equation}
    \dot{\alpha_i^v} = \left( \alpha_i^{\mathrm{loc,}v} - \alpha_i^v \right) \, \frac{v_i^{\mathrm{sig}}}{l h_i} \, ,
    \label{eqn:decay_art_visc}
\end{equation}
where we set $l = 4.0$ \citep{Beck_2015}, which specifies the decay length of the AV. We set an initial value $\alpha_i^v = 0.02$ in all our simulations unless a different initial value is specified (see Section~\ref{Intrinsic_visc_SPH}).

\subsection{Physical viscosity}

While AV is necessary for the correct behaviour of SPH in treating shocks with ideal fluids, the physical viscosity rules viscous fluids according to the Navier-Stokes equation. 

The physical viscosity implemented in \textsc{OpenGadget3} follows \cite{Sijacki_2006}. Taking into account the summation notation for repeated Greek indices, the viscous stress tensor (equation \ref{eqn:sigma}) is discretised as:
\begin{equation}
    \sigma_{\alpha \beta} \Big{|}_i = \eta \left( \left. \frac{\partial v_{\alpha}}{\partial x_{\beta}} \right|_i + \left. \frac{\partial v_{\beta}}{\partial x_{\alpha}} \right|_i - \frac{2}{3} \delta_{\alpha \beta} \left. \frac{\partial v_{\gamma}}{\partial x_{\gamma}} \right|_i \right) + \zeta \delta_{\alpha \beta} \left. \frac{\partial v_{\gamma}}{\partial x_{\gamma}} \right|_i \, .
\end{equation}

The implementation of the shear and bulk viscosity is made separately. The change in the acceleration due to the shear viscosity reads:
\begin{multline}
    \displaystyle\frac{\mathrm{d}v_{\alpha}}{\mathrm{d}t}\bigg{|}_{{\tiny{i, \mathrm{\textrm{ shear}}}}} = \displaystyle\sum\limits_{j=1}^N m_j \, \left[ \frac{\eta_i \, \sigma_{\alpha \beta}\big{|}_i}{\rho_i^2} \left( \nabla_i W_{ij}(r, h_i) \right) \Big{|}_{\beta} + \right. \\
    \left. + \frac{\eta_j \, \sigma_{\alpha \beta}\big{|}_j}{\rho_j^2} \left( \nabla_i W_{ij}(r, h_j) \right) \Big{|}_{\beta} \right] \, ,
\end{multline}
where the product of $\eta$ and $\mathbf{\sigma}$ gives the shear part of the viscous stress tensor. 

As for the bulk viscosity, it is calculated using
\begin{multline}
    \displaystyle\frac{\mathrm{d}v_{\alpha}}{\mathrm{d}t}\bigg{|}_{{\tiny{i, \mathrm{\textrm{ bulk}}}}} = \displaystyle\sum\limits_{j=1}^N m_j \left[ \frac{\zeta_i \nabla \cdot v_i}{\rho_i^2} \, \nabla_i W_{ij}(r, h_i) + \right. \\
    \left. + \frac{\zeta_j \nabla \cdot v_j}{\rho_j^2} \, \nabla_i W_{ij}(r, h_i) \right] \, .
\end{multline}

The friction due to viscosity causes an increase in the entropy, which is computed using the entropic function $A_i$ (see \citet[][]{Sijacki_2006} for details) as:
\begin{equation}
    \displaystyle\frac{\mathrm{d} A_i}{\mathrm{d}t}\bigg{|}_{{\mathrm{\textrm{\tiny{shear}}}}} = \frac{1}{2} \frac{\gamma - 1}{\rho_i^{\gamma - 1}} \frac{\eta_i}{\rho_i} \, \sigma_i^2
    \label{eqn:entropy_visc1}
\end{equation}
\begin{equation}
    \displaystyle\frac{\mathrm{d} A_i}{\mathrm{d}t}\bigg{|}_{{\mathrm{\textrm{\tiny{bulk}}}}} = \frac{\gamma - 1}{\rho_i^{\gamma - 1}} \frac{\zeta_i}{\rho_i} \, (\nabla \cdot v_i)^2 \, .
    \label{eqn:entropy_visc2}
\end{equation}
The additional variations to the internal energy read:
\begin{equation}
    \displaystyle\frac{\mathrm{d} u_i}{\mathrm{d}t}\bigg{|}_{{\mathrm{\textrm{\tiny{shear}}}}} = \frac{1}{2} \frac{\eta_i}{\rho_i} \, \sigma_i^2 \, ,
\end{equation}
\begin{equation}
    \displaystyle\frac{\mathrm{d} u_i}{\mathrm{d}t}\bigg{|}_{{\mathrm{\textrm{\tiny{bulk}}}}} = \frac{\zeta_i}{\rho_i} \, (\nabla \cdot v_i)^2 \, .
\end{equation}

Since the bulk viscosity only becomes relevant in presence of shocks, we set $\zeta=0$ in our set-up, and only take the shear viscosity term into account. 
In order to simulate fluids with different amounts of viscosity, we take fractions of the shear viscosity coefficient $\eta$ (equation \ref{eqn:shear_visc_coeff}). We assume a constant fluid temperature of $3\cdot10^7$ K. This is inside the range of ICM and circumgalactic-medium (CGM) temperatures, although might not be extended to the ISM regime.

\subsection{Kernel functions}
\label{Kernel functions}

Our set-up features a density step: this is the case of an irregular particle distribution. As a consequence, the so-called `E$_0$ error' \citep[e.g.][]{Read_2010, Hu_2014} may arise, caused by particles not being perfectly arranged according to an ordered pattern. This error is the attempt of SPH to restore the particle order and it can produce spurious results in fluid mixing \citep[e.g.][]{Hopkins_2013, Beck_2015, Wadsley_2017}. The `E$_0$ error' can be reduced by increasing $N_{\mathrm{ngb}}$. However, a high $N_{\mathrm{ngb}}$ leads to the `pairing instability' \citep[e.g.][]{Schuessler_1981, Price_2012} with a cubic spline kernel. 

The `pairing instability' is prone to kernel functions with a non-positive definite Fourier transform and it dominates the evolution of the system for large $N_{\mathrm{ngb}}$. \cite{Dehnen_2012} proposed a Wendland function as a kernel \citep{Wendland_1995} to address the `E$_0$ error' and the `pairing instability' simultaneously. In this work we employ a Wendland $C^6$ kernel in all our simulations corrected for bias in the central density. However, we point out that employing a higher $N_{\mathrm{ngb}}$ also implies a larger computational time (especially in MFM, where a larger set of Riemann problems has to be solved). 

\subsection{Set of simulations}

We have performed a total of 56 simulations using the code \textsc{OpenGadget3}, an improved version of \textsc{P-Gadget2/3} \citep{Springel_2005}, employing a modern SPH implementation \citep{Beck_2015} and a new MFM scheme (Groth et al, in prep). We used four different set-ups (see table \ref{tab:set_simulations} for a summary): 
\begin{itemize}
    \item Five simulations use SPH, the fiducial ICs described in Section~\ref{set-up}, and a constant AC to provide an ideal benchmark for comparison. In other five runs we do not introduce any initial perturbation in the $y$-velocity, while an additional run has no perturbations but includes a TDAC in order to measure the intrinsic viscosity and diffusion of the code. Moreover, we perform: eleven runs with different amounts of physical viscosity with initial perturbation, eleven simulations without initial perturbation with $N_{\mathrm{ngb}} = 150$ and eleven runs without initial perturbation using $N_{\mathrm{ngb}} = 295$. The runs with physical viscosity will provide us information on how viscosity affects the KHI and the measurement of the total viscosity of the code.
    \item Three runs adopt the fiducial KHI set-up using MFM and five other runs are carried out without initial perturbation for comparison with the runs with SPH.
    \item Two additional simulations with SPH using the ICs described in \citet{Read_2010} (see Section~\ref{Read_IC}). These runs allow us to explore how the simulation results are sensitive to the adopted ICs.
    \item Two additional runs with MFM and the ICs described in \citet{Read_2010} (see Section~\ref{Read_IC}) for comparison with the runs using the fiducial ICs with MFM. 
\end{itemize}

\begin{table*}
    \centering
    \caption{Description of all of simulations employed in this work.}
    \begin{tabular}{|c||c|c|c|c|c|c|c|}
        \hline
        Label & Code & Kernel & $N_{\mathrm{ngb}}$ & IC & AC & Phys. Visc. & Number of runs \\
        \hline \hline
        \multirow{5}{*}{OG-SPH} & \multirow{5}{*}{SPH} & \multirow{5}{*}{Wendland $C^6$} & $150-350$ & \citet{Murante_2011} & Constant & No & 5 \\
         &  &  & $150-350$ & No Perturbation & Constant & No & 5\\
         &  &  & $295$ & No Perturbation & Time Dependent & No & 1 \\
         &  &  & $295$ & \citet{Murante_2011} & Constant & $10^{-4}\,\eta - 10^{-2}\,\eta$ & 11 \\
         &  &  & $150, 295$ & No Perturbation & Constant & $10^{-4}\,\eta - 10^{-2}\,\eta$ & 22 \\
        \hline
        \multirow{2}{*}{OG-MFM} & \multirow{2}{*}{MFM} & \multirow{2}{*}{Wendland $C^6$} & $150, 200, 295$ & \citet{Murante_2011} &  & No & 3 \\
         &  &  & $150-350$ & No Perturbation &  & No & 5 \\
        \hline
        OG-SPH-Read & SPH & Wendland $C^6$ & $150, 295$ & \citet{Read_2010} & Constant & No & 2 \\
        \hline
        OG-MFM-Read & MFM & Wendland $C^6$ & $150, 295$ & \citet{Read_2010} &  & No & 2 \\
        \hline
    \end{tabular}
    \label{tab:set_simulations}
\end{table*}

\section{Results for SPH} \label{SPH_results}

\begin{figure*}
    \centering
	\includegraphics[width=\textwidth]{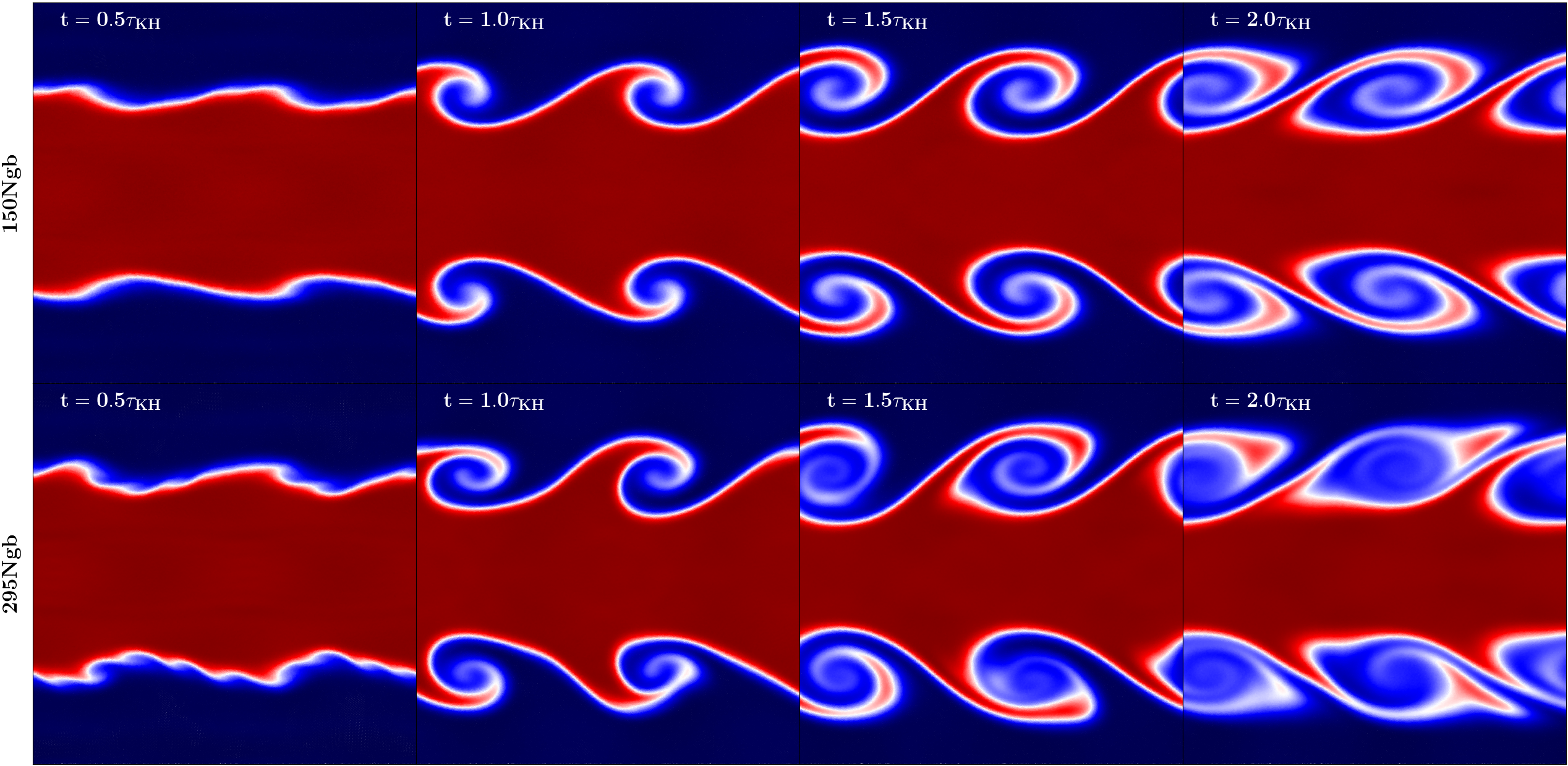}
    \caption{Projection of the mass weighted density for four different times (left to right) and two different $N_{\mathrm{ngb}}$, with $N_{\mathrm{ngb}} = 150$ (top row) and $N_{\mathrm{ngb}} = 295$ (bottom row), using OG-SPH. In both cases the instability can fully develop showing its characteristic roll. However, with $N_{\mathrm{ngb}} = 150$ the growth is slightly slower compared to the case with $N_{\mathrm{ngb}} = 295$.}
    \label{fig:colormaps_SPH}
\end{figure*}

In the first set of simulations we used OG-SPH and varied $N_{\mathrm{ngb}}$ to investigate the fluid mixing properties as a function of $N_{\mathrm{ngb}}$. Fig.~\ref{fig:colormaps_SPH} shows the colormaps of the density at different times for both $N_{\mathrm{ngb}} = 150$ and $N_{\mathrm{ngb}} = 295$. In a first qualitative approach, the colormaps show that the instability fully develops showing the characteristic roll of the KHI in both cases, although the growth with $N_{\mathrm{ngb}} = 150$ (top row) is slightly slower than $N_{\mathrm{ngb}} = 295$ (bottom row). In addition, at $t = 0.5 \tau_\mathrm{KH}$ with $N_{\mathrm{ngb}} = 295$ some secondary instabilities can be seen. These are caused by the contact discontinuity not being in perfect pressure equilibrium, which triggers a sound wave traveling across the volume. However, the secondary instabilities are successfully suppressed at later times, allowing only the main mode to grow. This leads to a highly symmetric result at later times.

\subsection{Growth of the KHI}
\subsubsection{Amplitude analysis} \label{amplitude_SPH}

Theory predicts that the initial perturbation triggers the instability and its amplitude starts to increase approximately linearly until it saturates with a height of $\sim \lambda/2$ \citep[see e.g.][]{Roediger_2013}. The amplitude is expected to  decrease afterwards, as each billow pairs with the adjacent ones \citep[e.g.][]{Rahmani_2014}. The height of the roll is a good indicator of how good the code captures the evolution of the KHI, and, conversely, which code suppresses more the instability and prevents its growth. The top panel of Fig.~\ref{fig:amplitude_McNally_SPH} shows the amplitude reached by the roll depending on $N_{\mathrm{ngb}}$ (see appendix \ref{app:amplitude} for a description of the method employed to compute the amplitude). 
\begin{figure}
	\includegraphics[width=\columnwidth]{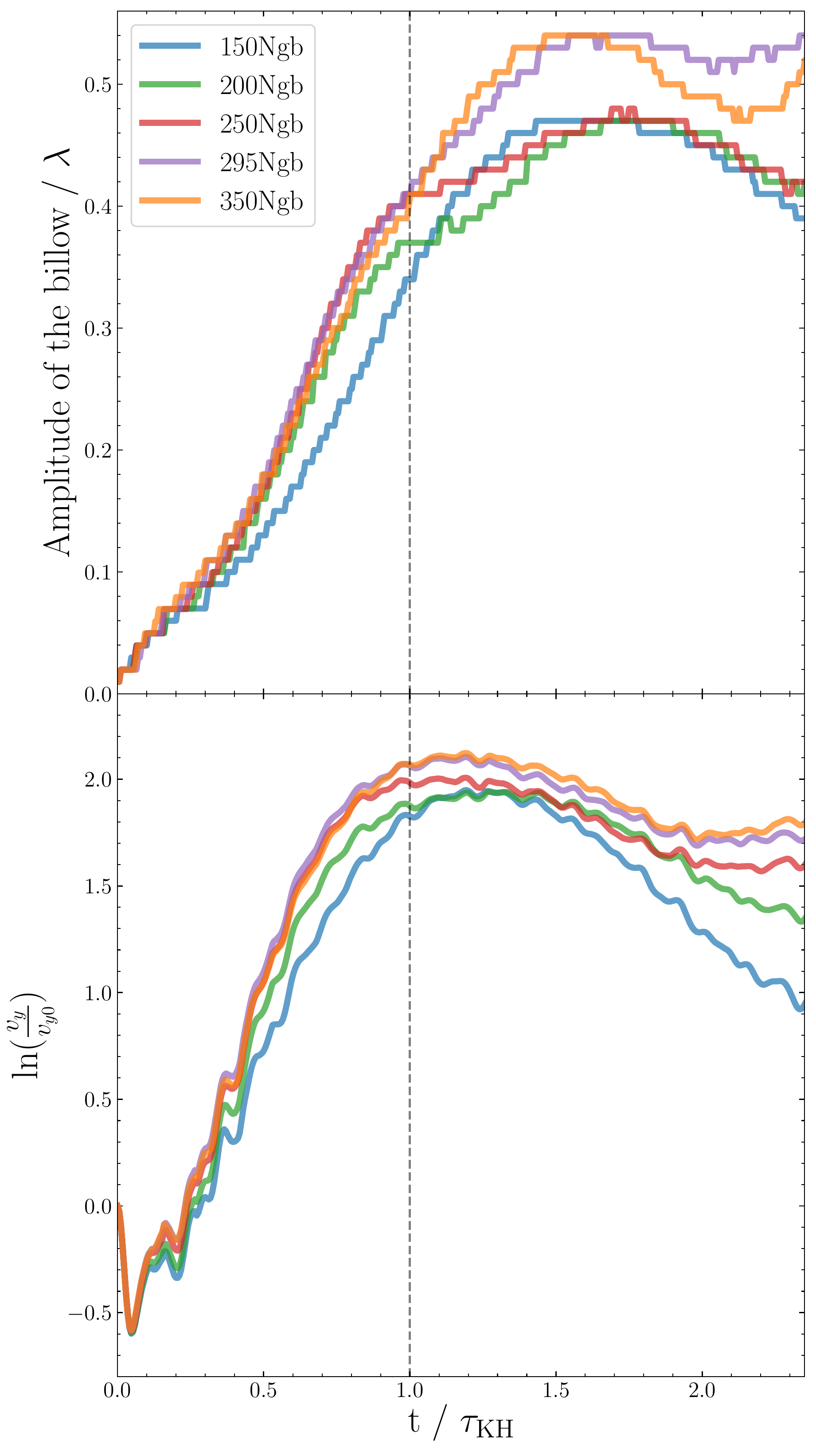}
    \caption{\textit{Top panel}: Temporal evolution of the height of the rolls for different $N_{\mathrm{ngb}}$ with OG-SPH. The amplitude of the rolls is dependent on $N_{\mathrm{ngb}}$, where the expected height of $\sim \lambda/2$ is not reached with a low $N_{\mathrm{ngb}}$, but it is reached for the runs with $N_{\mathrm{ngb}} > 250$. The instability reaches the maximum amplitude at times later than $\sim 1\tau_\mathrm{KH}$, which can be due to some intrinsic viscosity of the code. \textit{Bottom panel}: Change of amplitude of the $y$-velocity with time for the different $N_{\mathrm{ngb}}$ using OG-SPH. The run with $N_{\mathrm{ngb}} = 150$ shows a slower growth compared to the others and, as soon as we increase $N_{\mathrm{ngb}}$, the instability grows faster. This speed difference in the growth of the instability depending on $N_{\mathrm{ngb}}$ is the consequence of the intrinsic viscosity of the codes. The small decay at early times is due to the loss of kinetic energy of the particles by moving along the $y$ axis through a fluid flowing in the opposite direction.}
    \label{fig:amplitude_McNally_SPH}
\end{figure}

The height reached by the rolls depends on $N_{\mathrm{ngb}}$: for $N_{\mathrm{ngb}} \leq 250$, the amplitude reached is smaller than $0.5\lambda$; for $N_{\mathrm{ngb}} > 250$, the code is able to reach a height $> 0.5\lambda$. The plot also shows a general linear growth until approximately $t = \tau_\mathrm{KH}$ (vertical dashed line), which is what we expect for the ideal case. However, despite the simulations correspond to the ideal case, there is some intrinsic viscosity that delays the reach of the maximum amplitude. Each code has a different amount of numerical viscosity (see Section~\ref{Intrinsic_visc_SPH}) and this explains why the runs using a higher $N_{\mathrm{ngb}}$ evolve faster than the ones with a lower $N_{\mathrm{ngb}}$.

\subsubsection{Velocity analysis} \label{growth_vy_SPH}

The $y$-velocity is expected to grow exponentially until $t=\tau_\mathrm{KH}$ (see Section~\ref{KHI}). To measure the growth rate of the instability, depending on $N_{\mathrm{ngb}}$, we calculate how the $y$-velocity of the particles changes with time. This tells us how well a code can reproduce the KHI. To compute the amplitude of the $y$-velocity we use a discrete convolution of the sinusoidal perturbation \citep[see appendix \ref{app:v_y} for a detailed description of the method;][]{Sijacki_2006, Obergaulinger_2020}.

The results shown in the bottom panel of Fig.~\ref{fig:amplitude_McNally_SPH} exhibit a general linear trend (note that the $y$ axis is in log scale) with an initial decrease of the amplitude of the velocities. This is due to the loss of kinetic energy of the particles by moving along the $y$ axis through a fluid streaming in the opposite direction \citep[e.g.][]{Junk_2010}. The more viscous a fluid is, the more significant is the loss of kinetic energy.

The results show a correlation between $N_{\mathrm{ngb}}$ and the growth rate of the $y$-velocity, with the simulation with $N_{\mathrm{ngb}} = 150$ growing the slowest. This analysis also agrees with the results observed in the growth of the height of the roll during the linear phase, where the amplitude of the run with $N_{\mathrm{ngb}} = 150$ evolved the slowest, followed by the case with $N_{\mathrm{ngb}} = 200$. The run with $N_{\mathrm{ngb}} = 250$ has a similar $y$-velocity at early times compared to the cases with $N_{\mathrm{ngb}} = 295$ and $N_{\mathrm{ngb}} = 350$. However, the maximum velocity reached is lower than the runs with higher $N_{\mathrm{ngb}}$, which explains why the height of the rolls also follows a similar trend at the beginning, but the maximum amplitude reached is smaller. Such a difference in the growth rate can be also explained in terms of the numerical viscosity of each code (see Section~\ref{Intrinsic_visc_SPH}). 

\subsection{Diffusion} \label{diffusion_SPH}

An important consequence of the KHI is the mixing process between the two fluids. Diffusion of thermal energy is the main mixing process at early times due to the fact that the roll has not formed yet. The mixing comes from the movement of the particles and redistributes energy from regions with high specific internal energy to regions with low specific internal energy. In the case of SPH, it must be added artificially via the artificial conductivity, which makes it difficult to identify whether the amount added is too high or too low. For that purpose, in this section, we analyse how diffusive is Gadget's SPH-solver, depending on $N_{\mathrm{ngb}}$ with a constant AC and a time dependent AC (TDAC). 

Since AC widens the fluid interface by smoothing the discontinuity due to the random movement of the particles, we analyse the thickness of the interface at early times to measure the amount of diffusion of the code (as described in appendix \ref{app:diffusion}). For this analysis, we run the simulations again without adding any initial perturbation. Despite no initial perturbation is added, some small scale instabilities can be triggered numerically, affecting our measurement. To avoid that, we compute the diffusion only until $t = 0.4\tau_{\mathrm{KH}}$.
\begin{figure}
	\includegraphics[width=\columnwidth]{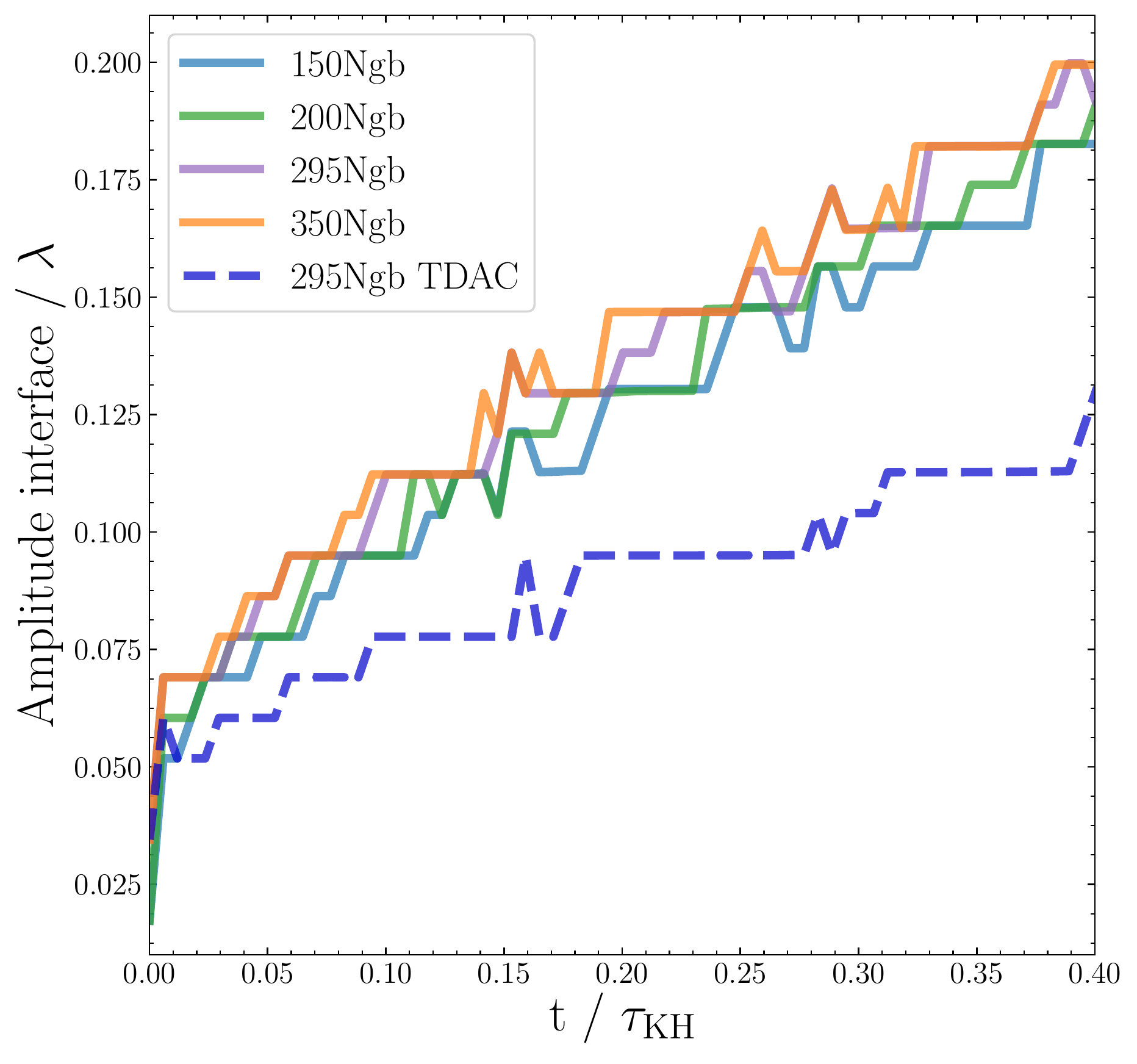}
    \caption{Measurement of diffusion for the different runs at early times. The simulations with a constant AC (solid lines) happen to be excessively diffusive compared to the ones with a TDAC (dash-dotted line), which keeps the diffusion low at early times. This might produce an excess of mixing in the runs with a constant AC in the long term evolution.}
    \label{fig:Diffusion_SPH}
\end{figure}

Fig.~\ref{fig:Diffusion_SPH} shows the evolution of the amplitude of the interface: the diffusion of the code is independent of $N_{\mathrm{ngb}}$ when a constant AC is used. In contrast, the results obtained with a TDAC show that, when the AC is not constant, a much lower diffusive state can still enable the growth of the instability. The TDAC reduces the diffusion of the code to the minimum value needed to reproduce the instability. This suggests that the constant AC added artificially to the code is higher than the one needed and produces more diffusive results than the one expected. As a result, there might be an excess of mixing between the two fluids in the long term evolution when constant AC is applied throughout the simulation.

\subsection{Intrinsic viscosity} \label{Intrinsic_visc_SPH}

The results shown in sections \ref{amplitude_SPH} and \ref{growth_vy_SPH} reveal that each code has an intrinsic viscosity depending on $N_{\mathrm{ngb}}$. For a lower $N_{\mathrm{ngb}}$ this intrinsic viscosity produces a slower growth of the instability and a lower height of the rolls. 

\begin{figure}
	\includegraphics[width=\columnwidth]{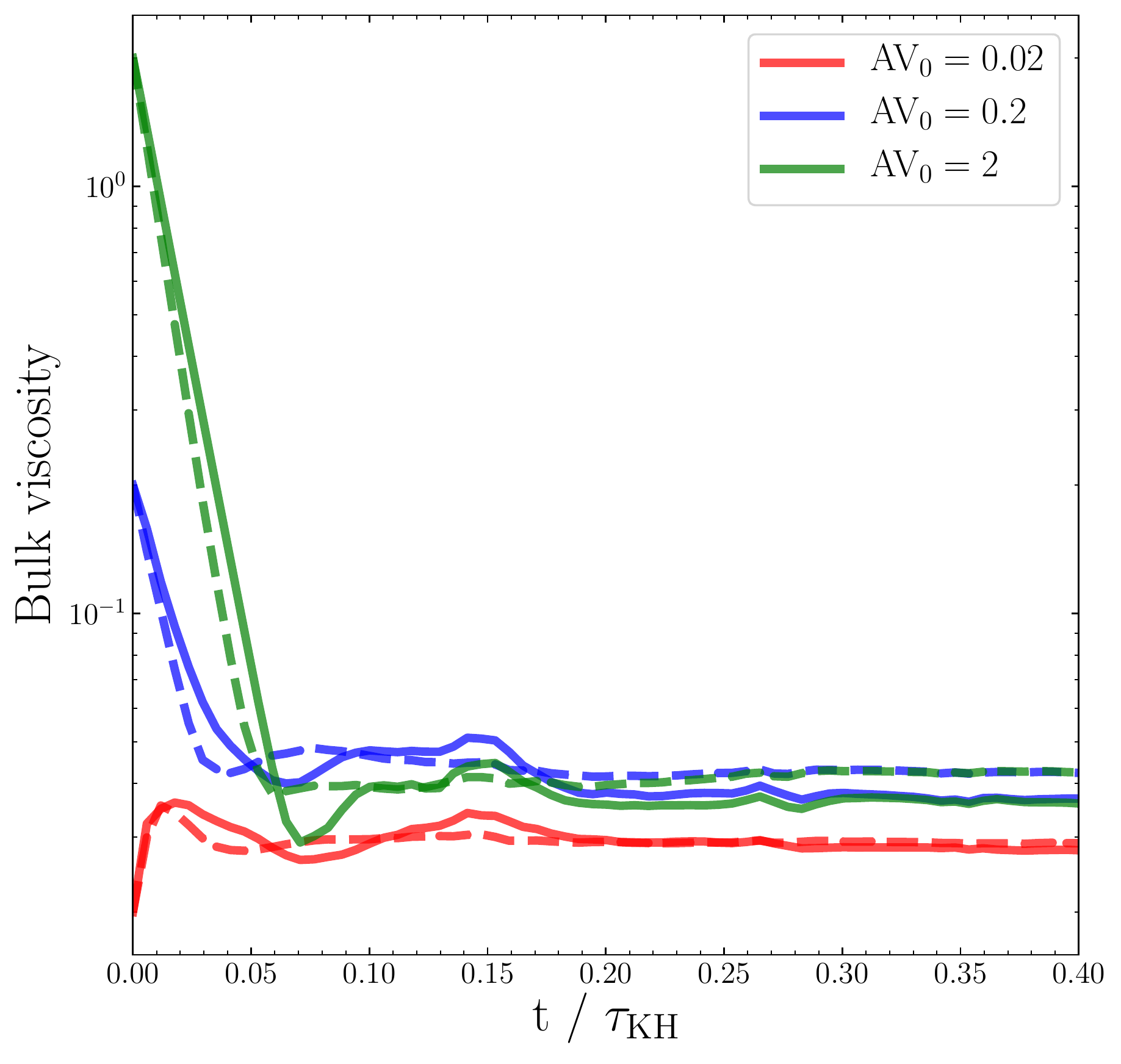}
    \caption{Change of the average AV with time in the whole simulation domain for three different initial values for the runs with $N_{\mathrm{ngb}} = 150$ (dashed lines) and $N_{\mathrm{ngb}} = 295$ (solid lines). The AV is successfully reduced at early times, showing that it does not affect our results. The fact that the runs with $N_{\mathrm{ngb}} = 150$ decrease slightly faster is because the decay of the AV depends on the smoothing length of the kernel (see equation \ref{eqn:decay_art_visc}), which depends on $N_{\mathrm{ngb}}$.}
    \label{fig:Art_visc_SPH}
\end{figure}

Fig.~\ref{fig:Art_visc_SPH} shows that the AV is successfully reduced at early times independently of the initial value we set. 
This means that the intrinsic viscosity observed does not arise from the AV added in OG-SPH, but it is intrinsic to the code and depends on $N_{\mathrm{ngb}}$. For a more quantitative analysis, we measured the intrinsic viscosity of each code following the method explained in appendix \ref{app:viscosity} and obtained the results shown in table \ref{tab:intrinsic_visc} and Fig.~\ref{fig:Intrinsic_visc}. These results show that, in the case of OG-SPH, the run with $N_{\mathrm{ngb}} = 150$ has the largest amount of intrinsic viscosity and, as soon as $N_{\mathrm{ngb}}$ is increased, the intrinsic viscosity is reduced. The runs with $N_{\mathrm{ngb}} = 250$, $N_{\mathrm{ngb}} = 295$ and $N_{\mathrm{ngb}} = 350$ have a very similar viscosity, which could explain the similar rate of growth measured and shown in the bottom panel of Fig.~\ref{fig:amplitude_McNally_SPH}. However, the overall intrinsic viscosity of OG-SPH remains low in all cases and, despite some influence in the development of the KHI, it does not suppress its growth.

\section{Results for MFM} \label{MFM_results}

\begin{figure*}
    \centering
	\includegraphics[width=\textwidth]{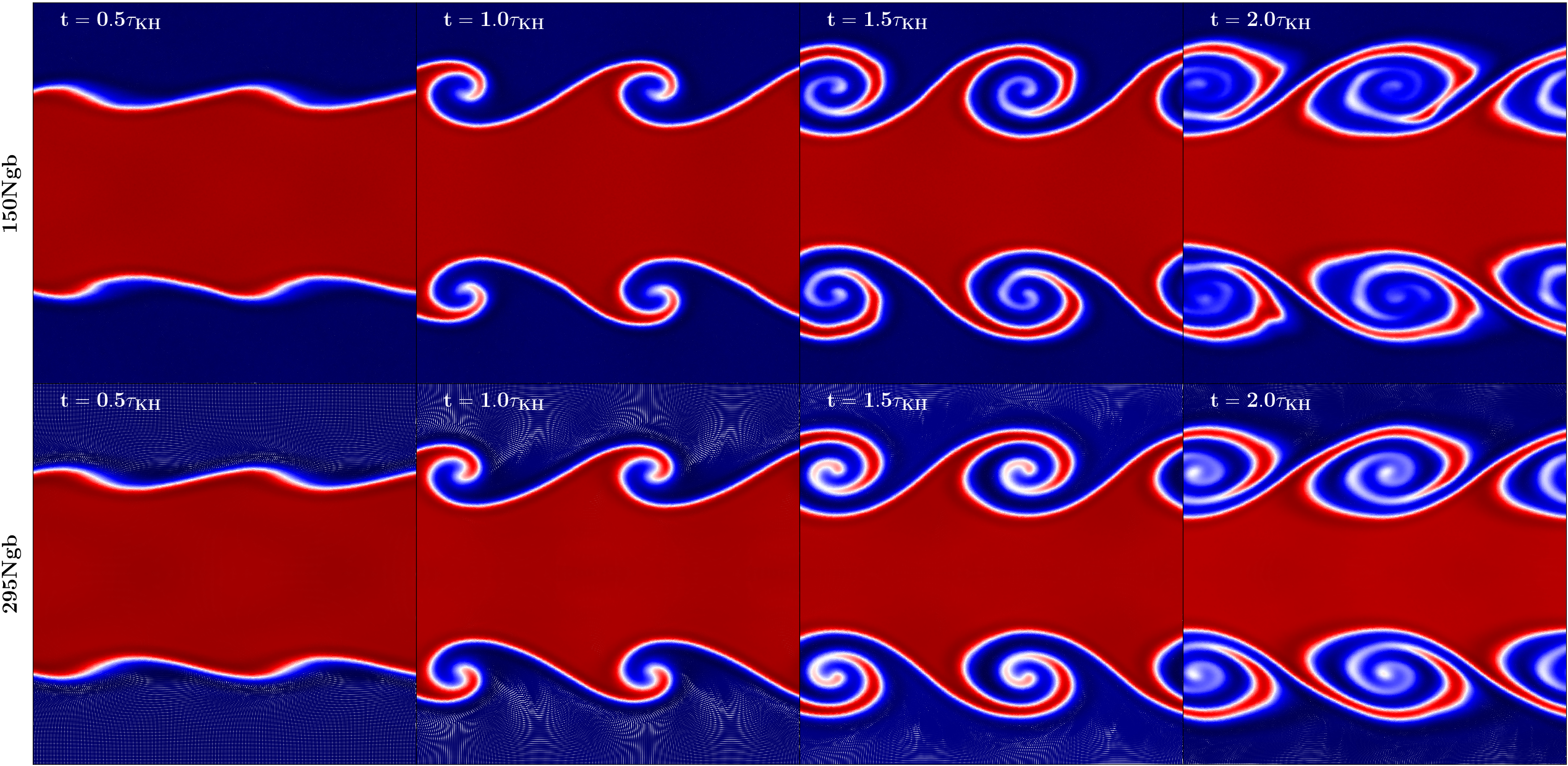}
    \caption{Same plot as in Fig.~\ref{fig:colormaps_SPH} but using OG-MFM in this case. Despite $N_{\mathrm{ngb}}$ is the same as the one we used in OG-SPH, the results of the two runs are more similar. Some secondary instabilities can be seen growing with $N_{\mathrm{ngb}} = 150$ at $t=2\tau_\mathrm{KH}$, but they do not lead to the breakdown of the billow. Also, the billows in this case are much less diffusive compared to OG-SPH.}
    \label{fig:colormaps_MFM}
\end{figure*}

We also tested the MFM scheme \citep[e.g.][]{Gaburov_2011, Hopkins_2015, Hubber_2018} implemented in our code \textsc{OpenGadget3}, with the same set-up as OG-SPH using different $N_{\mathrm{ngb}}$. Details on our MFM implementation are presented in Groth et al. (2022, in prep). The `E$_0$ error' problem does not occur in MFM-like schemes, so we should be able to get satisfactory results with a low $N_{\mathrm{ngb}}$ and, in principle, there is no need of using a Wendland kernel. That is why the cubic spline kernel is widely used with MFM in cosmological simulations \citep[e.g.][]{Dave_2016, Hopkins_2018, Rennehan_2021}. However, we found that with a cubic spline kernel the secondary instabilities are not successfully suppressed, allowing their growth and provoking the breakdown of the billow (see appendix \ref{cubic_spline}). Because of this, we preferred to use a Wendland $C^6$ kernel in our simulations with OG-MFM. 

In Fig.~\ref{fig:colormaps_MFM} we show the column density for $N_{\mathrm{ngb}} = 150$ and $N_{\mathrm{ngb}} = 295$. The rolls can successfully grow in both cases leading to a symmetric system at late times. Both shapes are similar and, despite the runs with $N_{\mathrm{ngb}} = 150$ show some secondary instabilities in the inner parts of the roll, they do not introduce the breakdown of the billow. The spiral of the billows is clearly visible, indicating a lower diffusion compared to OG-SPH.

\subsection{Growth of the KHI}
\subsubsection{Amplitude analysis} \label{Amplitude_MFM}

\begin{figure}
	\includegraphics[width=\columnwidth]{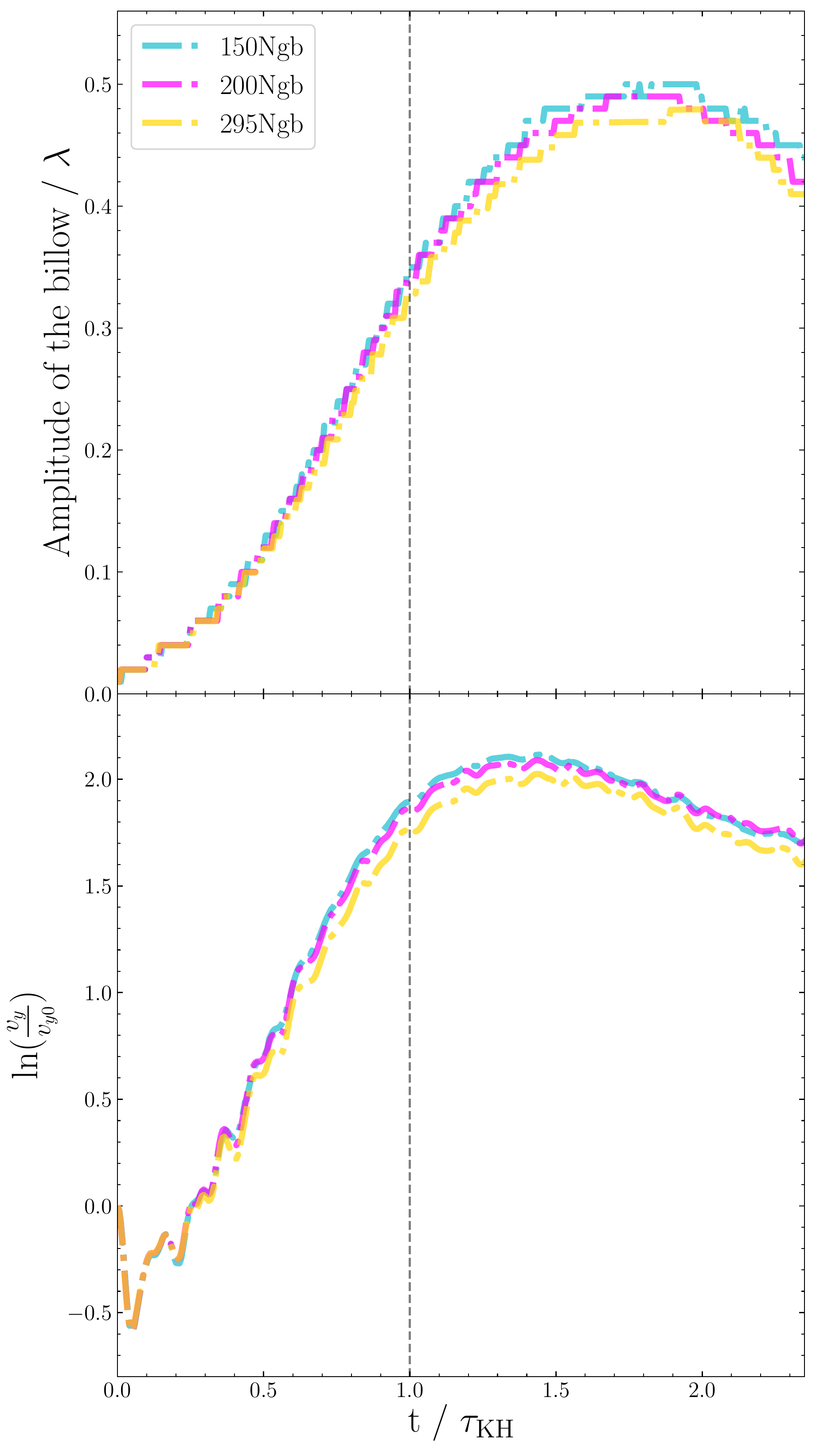}
    \caption{\textit{Top panel}: Evolution of the height of the billows depending on $N_{\mathrm{ngb}}$ with OG-MFM. The amplitude of the rolls does not depend on $N_{\mathrm{ngb}}$ and the three runs perform similarly. All the runs reach an amplitude of $\sim \lambda/2$. However, the maximum is reached later compared to OG-SPH. \textit{Bottom panel}: Growth of the $y$-velocity in OG-MFM depending on $N_{\mathrm{ngb}}$. The three runs have a very similar behaviour. If we compare these results with the ones we got with OG-SPH (bottom panel of Fig.~\ref{fig:amplitude_McNally_SPH}), the results with OG-MFM are less steep than OG-SPH, which means that the instability grows slower and can explain that the maximum in the amplitude is reached later.}
    \label{fig:amplitude_McNally_MFM}
\end{figure}

We perform the same analysis done for OG-SPH for a reliable comparison between OG-SPH and OG-MFM. In the top panel of Fig.~\ref{fig:amplitude_McNally_MFM} we show how the amplitude changes with time for the cases with $N_{\mathrm{ngb}} = 150$, $N_{\mathrm{ngb}} = 200$ and $N_{\mathrm{ngb}} = 295$. The growth of the three instabilities is very similar, reaching a maximum amplitude close to $\sim \lambda/2$. The total height reached is slightly lower for a higher $N_{\mathrm{ngb}}$. However, the difference between the three runs is overall negligible. In comparison with the results with OG-SPH (see top panel of Fig.~\ref{fig:amplitude_McNally_SPH}), the peak is reached at later times in OG-MFM, which suggests a larger amount of intrinsic viscosity of OG-MFM.

\subsubsection{Velocity analysis} \label{McNally_MFM}

The results obtained when analysing the growth rate of the instability (see bottom panel of Fig.~\ref{fig:amplitude_McNally_MFM}) show a very similar behaviour independently of $N_{\mathrm{ngb}}$. By comparing with OG-SPH (bottom panel of Fig.~\ref{fig:amplitude_McNally_SPH}), the slope is steeper in OG-SPH, meaning that the KHI evolves faster in OG-SPH than in OG-MFM. The reason for this is likely the intrinsic viscosity of the code and could explain why the maximum peak is reached later in OG-MFM than in OG-SPH.

\subsection{Diffusion}

\begin{figure}
	\includegraphics[width=\columnwidth]{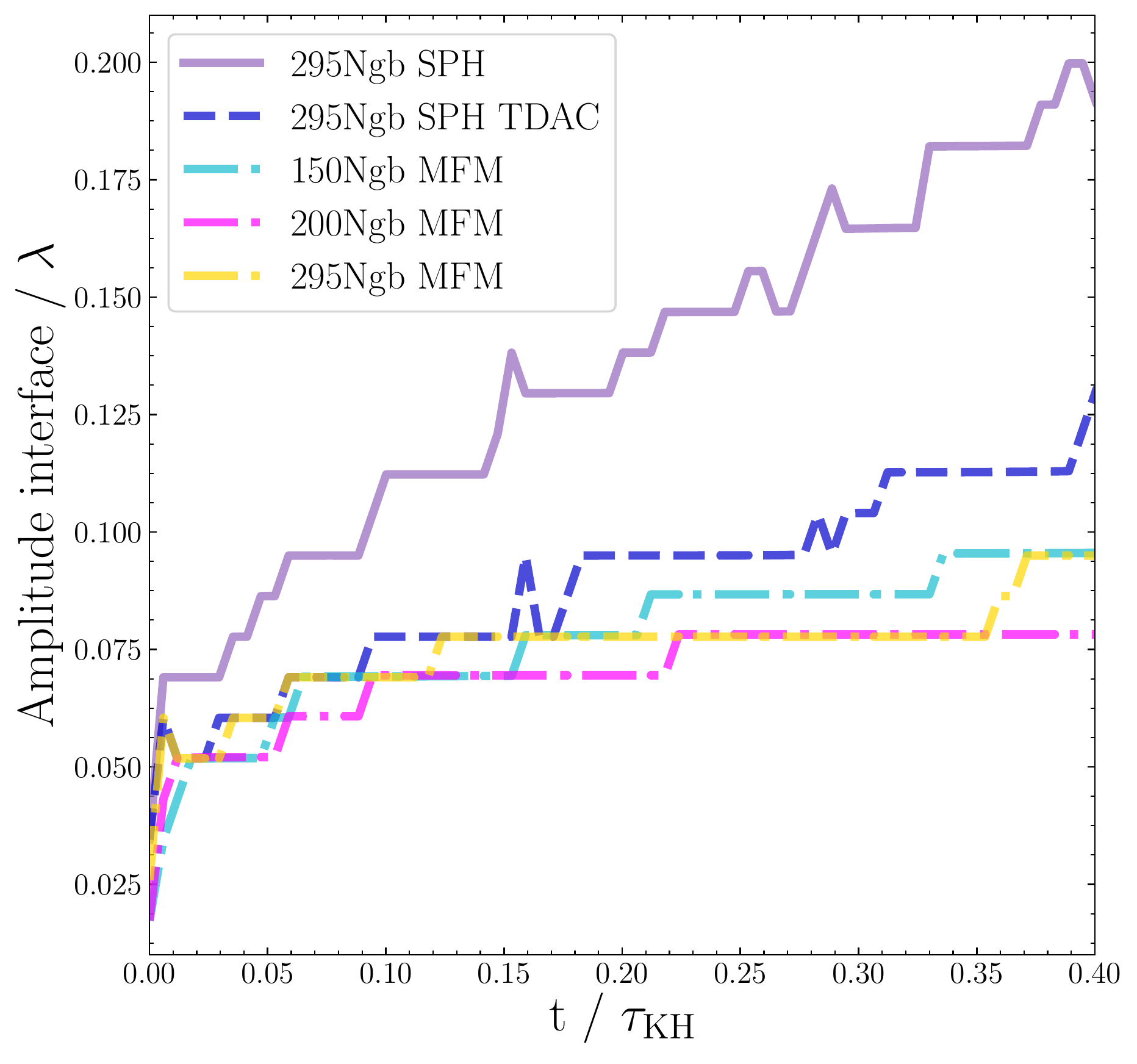}
    \caption{The runs with OG-MFM are less diffusive at early times than OG-SPH due to the fact that the diffusion is not added artificially, but it is intrinsic to the code. However, the run with TDAC follows a similar trend to OG-MFM.}
    \label{fig:Diffusion_MFM}
\end{figure}

In classic SPH methods, artificial diffusion terms are necessary to enable fluid mixing. However, in Godunov-type Riemann based methods, such as MFM, diffusion is naturally added as a resolution dependent mechanism due to the formulation of the equations of motion in a finite volume (FV) scheme. This procedure typically circumvents the introduction of artificial diffusion terms to stabilize the scheme numerically. Although diffusion in mesh codes is purely numerical due to the advection error, its value is close to the expected physical diffusion when the velocity of the fluids is low \citep[e.g.][]{Wadsley_2017}. To this end, we plot the results in Fig.~\ref{fig:Diffusion_MFM} together with the diffusion measured for OG-SPH with a constant AC and with a time dependent AC.

The results show that the OG-MFM runs are less diffusive than the runs with OG-SPH and constant AC. This confirms the statement made in Section~\ref{diffusion_SPH} that a constant AC introduces too much diffusion in our results. However, the diffusion obtained with the TDAC run is comparable to the ones obtained with OG-MFM. This means that the TDAC indeed reduces the artificial conductivity to the minimum value needed, avoiding over-mixing.

\subsection{Intrinsic viscosity}

\begin{figure}
	\includegraphics[width=\columnwidth]{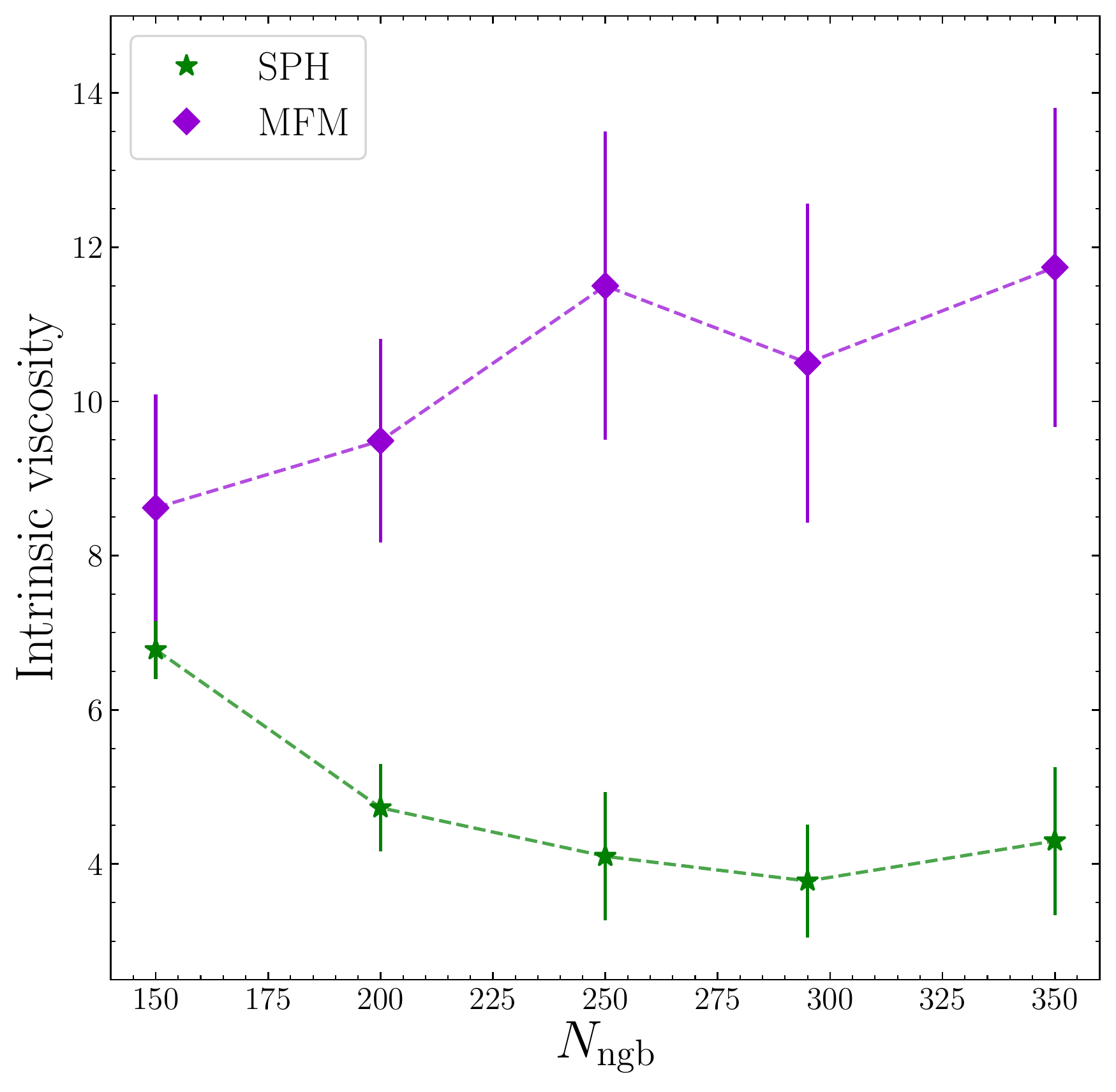}
    \caption{Amount of intrinsic viscosity depending on $N_{\mathrm{ngb}}$. The amount of intrinsic viscosity seems to decrease with $N_{\mathrm{ngb}}$ in OG-SPH. Whereas, using OG-MFM, it tends to increase with $N_{\mathrm{ngb}}$, reaching a larger intrinsic viscosity than OG-SPH for a high $N_{\mathrm{ngb}}$.}
    \label{fig:Intrinsic_visc}
\end{figure}

Although we are simulating ideal fluids, as we analysed in Section~\ref{Intrinsic_visc_SPH} for the case of OG-SPH, hydro solvers happen to have some intrinsic viscosity that can affect the results of our simulations. In the case of OG-SPH this viscosity depended on $N_{\mathrm{ngb}}$, finding that the case with $N_{\mathrm{ngb}} = 150$ produces the most viscous fluids. In this section we compute the intrinsic viscosity of OG-MFM for the different $N_{\mathrm{ngb}}$ in order to compare the results with the ones obtained with OG-SPH. We show the results of the intrinsic viscosity obtained from OG-MFM together with the ones from OG-SPH in table \ref{tab:intrinsic_visc} and plot them in Fig.~\ref{fig:Intrinsic_visc}.
In all the cases, the fluids simulated with OG-MFM are more viscous compared to the ones simulated with OG-SPH. This could explain the differences observed between the results with OG-SPH and OG-MFM analysed in sections \ref{Amplitude_MFM} and \ref{McNally_MFM}. However, despite the analysis reveals a larger amount of intrinsic, numerical viscosity in OG-MFM compared to OG-SPH, this excess of intrinsic viscosity does not inhibit the growth of the KHI and the rolls can properly develop. In contrast to the flat trend of OG-SPH, the intrinsic viscosity of OG-MFM tends to increase when a higher $N_{\mathrm{ngb}}$ is employed. The amount of intrinsic viscosity of the code might vary depending on the slope limiter employed. In our simulations we have used the slope limiter suggested by \citet{Springel_2010} for the moving mesh code \textsc{AREPO}.

\section{Effect of Physical viscosity} \label{Physical_viscosity}

The viscosity of a fluid can determine its properties and behaviour, so in order to study this effect, we introduce viscosity in the system to quantify how previous results change when the assumption of inviscid fluid does not hold anymore. For this purpose, we use the OG-SPH code with $N_{\mathrm{ngb}} = 295$ to study the behaviour of the system depending on the amount of viscosity implemented. Since physical viscosity has not been implemented yet in our OG-MFM code, we analyse the behaviour of viscous fluids using OG-SPH only.

In order to study the effect that physical viscosity has on our results, we run 11 different simulations with 11 different fractions of Braginskii viscosity\footnote{As mentioned before, for the viscosity we will assume a constant temperature of the fluids of $3\cdot10^7$ K to match the typical conditions within the ICM.}. Table \ref{tab:viscosity} shows the 11 different fractions with the corresponding value of the dynamic viscosity in internal units, as well as the corresponding value of kinematic viscosity ($\nu$). Taking fractions of viscosity mimics the effect that magnetic field has in viscosity, which is suppressed in the direction of the magnetic field.

\begin{table*}
    \centering
    \caption{Different amounts of viscosity employed in our simulations with the actual viscosity computed for both $N_{\mathrm{ngb}} = 150$ and $N_{\mathrm{ngb}} = 295$ and their deviation with respect to the theoretical value.}
    \begin{tabular}{|c|c|c|c|c|c|c|}
        \hline
        Fraction & $\hat{\eta}$ & $\nu$ & Actual $\nu$ & Deviation & Actual $\nu$ & Deviation \\
        ($\eta$) &   &   & ($N_{\mathrm{ngb}} = 150$) & ($N_{\mathrm{ngb}} = 150$) &  ($N_{\mathrm{ngb}} = 295$) &  ($N_{\mathrm{ngb}} = 295$) \\
        \hline \hline
        $10^{-4}$ & $1.379\cdot10^{-7}$ & $3.305$ & $9.57\pm0.15$ & $6.26$ ($189.53\%$) & $5.69\pm0.31$ & $2.38$ ($72.15\%$) \\
        \hline
        $2.5\cdot10^{-4}$ & $3.449\cdot10^{-7}$ & $8.263$ & $14.68\pm0.38$ & $6.42$ ($77.65\%$) & $10.15\pm0.21$ & $1.89$ ($22.83\%$) \\
        \hline
        $5\cdot10^{-4}$ & $6.897\cdot10^{-7}$ & $16.527$ & $23.41\pm0.65$ & $6.88$ ($41.65\%$) & $18.20\pm0.32$ & $1.67$ ($10.13\%$) \\
        \hline
        $7.5\cdot10^{-4}$ & $1.035\cdot10^{-6}$ & $24.790$ & $31.46\pm0.97$ & $6.67$ ($26.91\%$) & $26.32\pm0.34$ & $1.53$ ($6.17\%$) \\
        \hline
        $10^{-3}$ & $1.379\cdot10^{-6}$ & $33.053$ & $39.25\pm0.80$ & $6.20$ ($18.75\%$) & $34.37\pm0.36$ & $1.25$ ($3.77\%$) \\
        \hline 
        $1.5\cdot10^{-3}$ & $2.069\cdot10^{-6}$ & $49.580$ & $55.27\pm0.76$ & $5.69$ ($11.48\%$) & $50.10\pm0.55$ & $0.52$ ($1.05\%$) \\
        \hline
        $2\cdot10^{-3}$ & $2.759\cdot10^{-6}$ & $66.106$ & $70.66\pm1.64$ & $4.55$ ($6.89\%$) & $66.50\pm0.56$ & $0.39$ ($0.60\%$) \\
        \hline
        $2.5\cdot10^{-3}$ & $3.449\cdot10^{-6}$ & $82.633$ & $87.34\pm1.28$ & $4.71$ ($5.70\%$) & $83.08\pm1.47$ & $0.45$ ($0.54\%$) \\
        \hline
        $5\cdot10^{-3}$ & $6.897\cdot10^{-6}$ & $165.265$ & $169.3\pm2.9$ & $4.07$ ($2.46\%$) & $168.0\pm4.1$ & $2.76$ ($1.67\%$) \\
        \hline
        $7.5\cdot10^{-3}$ & $1.035\cdot10^{-5}$ & $247.898$ & $249.9\pm9.2$ & $2.00$ ($0.81\%$) & $247.2\pm9.0$ & $-0.70$ ($0.28\%$) \\
        \hline 
        $10^{-2}$ & $1.379\cdot10^{-5}$ & $330.530$ & $326.9\pm3.6$ & $-3.64$ ($1.10\%$) & $334.0\pm5.5$ & $3.48$ ($1.05\%$) \\
        \hline
    \end{tabular}
    \label{tab:viscosity}
\end{table*}

Note that the kinematic viscosity is defined as
\begin{equation}
    \nu = \frac{\eta}{\rho} \, ,
\end{equation}
which means that it depends on the density of the fluid. Since we have two different fluids but the dynamic viscosity is the same for both of them, in our analysis we have computed the kinematic viscosity using the mean value of the kinematic viscosities of each fluid. 

\begin{figure*}
    \centering
	\includegraphics[width=\textwidth]{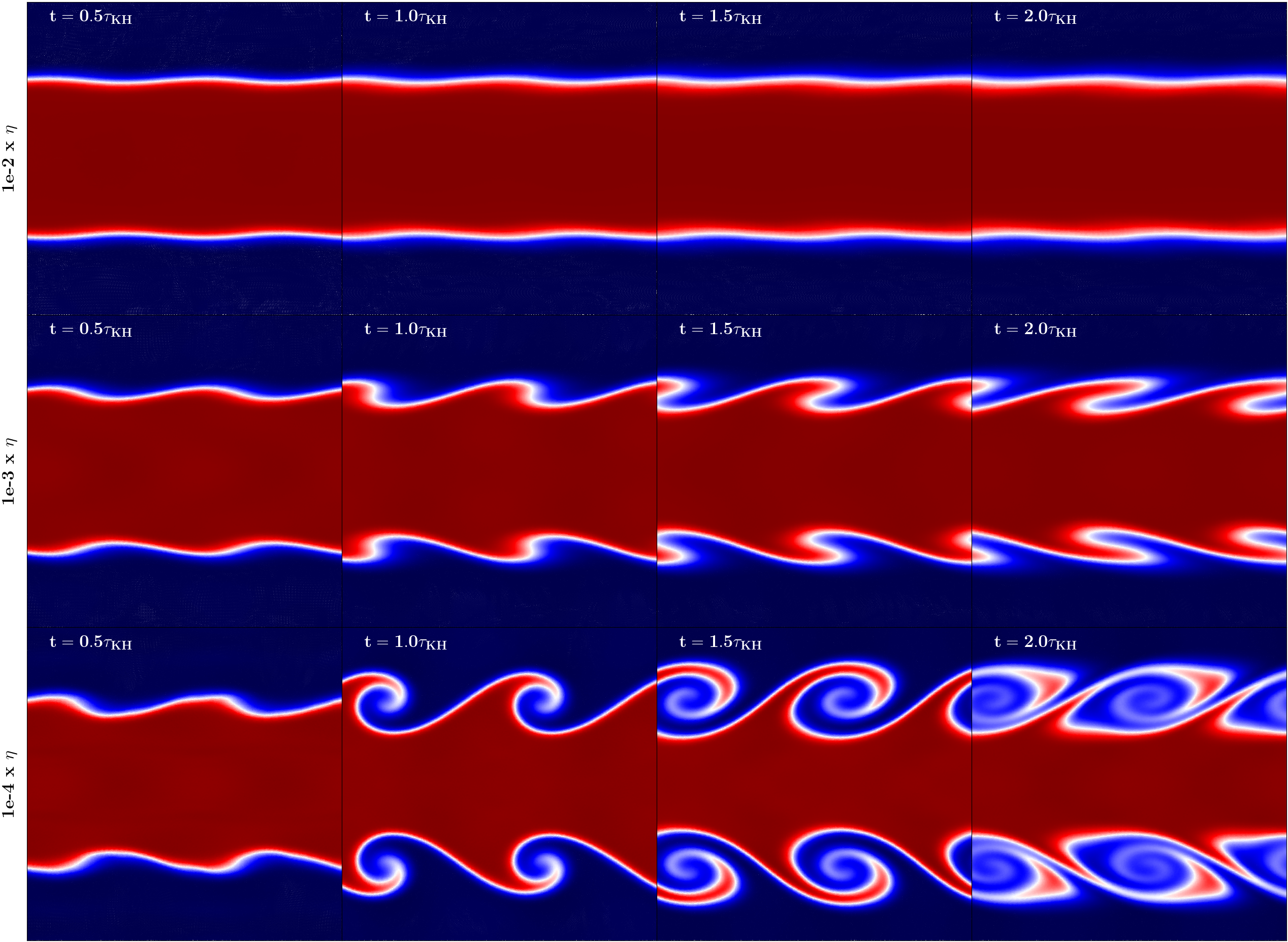}
    \caption{Projection of the mass weighted density for four different times (left to right) and three different values of the physical viscosity from higher viscosity (top row) to lower viscosity (bottom row) using OG-SPH. The run with the highest viscosity $10^{-2} \, \eta$ (top row) suppresses completely the growth of the KHI. In the case of $10^{-3} \, \eta$ (middle row) the instability is partially suppressed but there is still some growth of the perturbation. With a low viscosity $10^{-4} \, \eta$ (bottom row) the instability can grow properly showing similar results to the ideal case.}
    \label{fig:colormaps_visc}
\end{figure*}

As a first qualitative result, Fig.~\ref{fig:colormaps_visc} shows the mass weighted projected density for three different viscosities, $10^{-2} \, \eta$ (top), $10^{-3} \, \eta$ (centre) and $10^{-4} \, \eta$ (bottom). With a large amount of viscosity the instability is fully suppressed and the roll is no longer developed, but as soon as we decrease the amount of viscosity, the instability appears and grows with higher amplitude. In the particular case with a viscosity of $10^{-4} \, \eta$, the shape of the roll happens to be similar to the ones we got for the ideal case (see Fig.~\ref{fig:colormaps_SPH}).

\subsection{Growth of the KHI}
\subsubsection{Amplitude analysis}

\begin{figure}
	\includegraphics[width=\columnwidth]{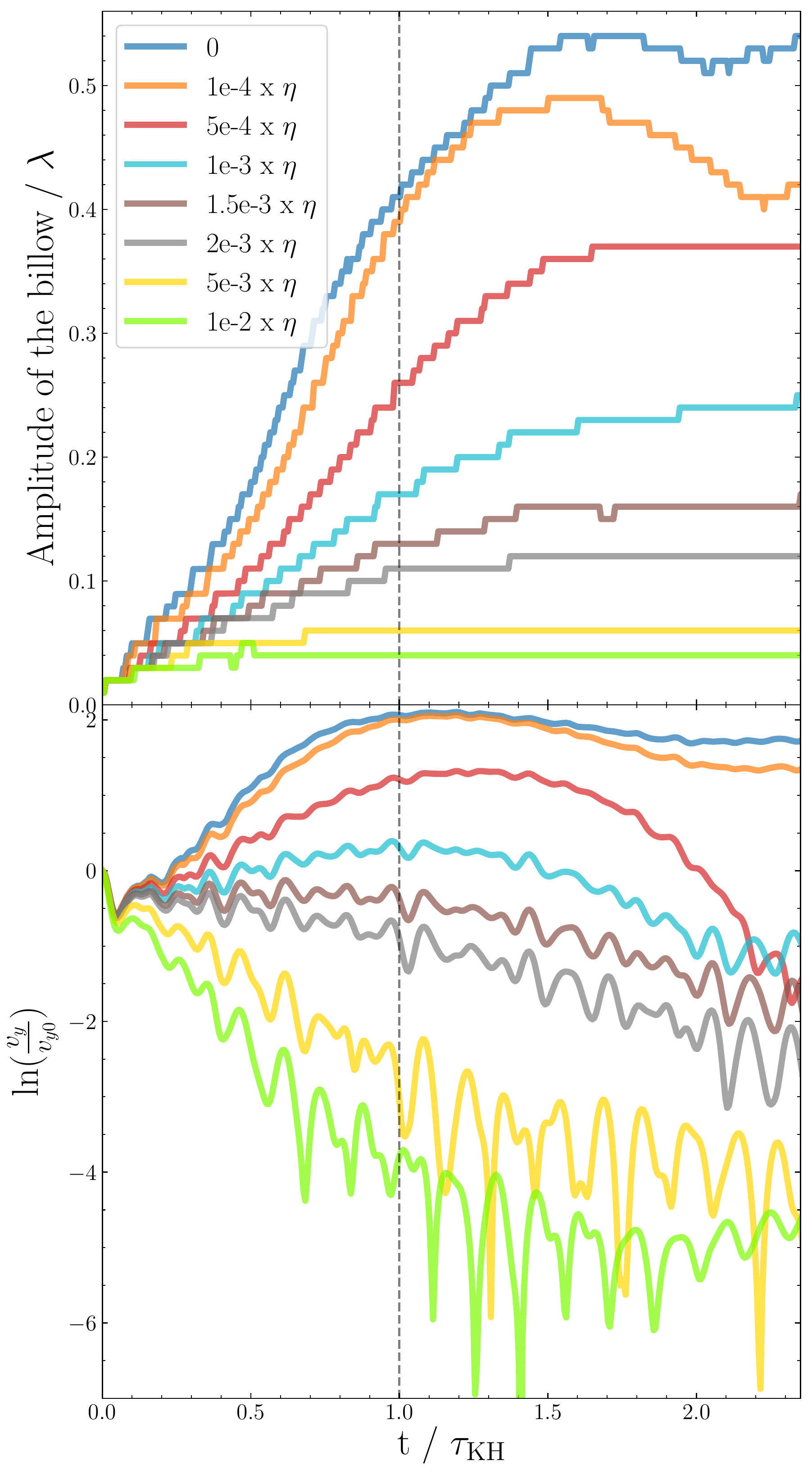}
    \caption{\textit{Top panel}: Height of the rolls depending on how viscous the fluids are. For the most viscous case ($10^{-2} \, \eta$), the amplitude barely increases, but as soon as we decrease the viscosity of the system, the rolls reach a higher amplitude until we have $10^{-4}\eta$, where the rolls reach an amplitude close to $\lambda/2$. However, this height is smaller than the one reached in the ideal case with OG-SPH and $N_{\mathrm{ngb}} = 295$. \textit{Bottom panel}: Evolution of the $y$-velocity depending on the amount of viscosity of the system. In a very viscous system ($10^{-2} \, \eta$) the KHI decays exponentially, while the case with the lowest amount of viscosity ($10^{-4} \, \eta$) follows a similar path to the ideal case. The amount of viscosity that produces no growth of the $y$-velocity (slope zero) characterizes a viscosity threshold above which the KHI is fully suppressed. This threshold can be determined numerically and compared to a theoretical estimate.}
    \label{fig:amplitude_McNally_visc}
\end{figure}

In order to study the effect of physical viscosity in detail, we perform the same analysis we did for the ideal case with OG-SPH. By measuring the height of the rolls we are able to study the level of suppression that the KHI suffers depending on how viscous the system is. The top panel of Fig.~\ref{fig:amplitude_McNally_visc} shows that the amplitude is reduced when the fluids are more viscous due to the fact that the friction between particles reduces the kinetic energy causing the KHI not to develop. This means that the instability is fully suppressed for the simulations with the highest viscosities, and as soon as the fluids are less viscous, the instability is able to develop with larger amplitudes. At early times we can see some increase of the amplitude due to the mixing of the fluids triggered by the thermal conduction. However, after this, the amplitude does not grow anymore. Despite the case with $10^{-4}\,\eta$ reaches an amplitude of almost $\lambda/2$, we can observe a difference compared to the ideal case with OG-SPH and $N_{\mathrm{ngb}} = 295$, where the maximum height reached is larger than the viscous case. 

\subsubsection{Velocity analysis}

To analyse the growth of the KHI we compute the $y$-velocity as we did in Section~\ref{growth_vy_SPH} and show the results in the bottom panel of Fig.~\ref{fig:amplitude_McNally_visc}. Due to friction, the more viscous the system is, the slower the KHI grows. For higher amounts of viscosity, the rate of the growth of the $y$-velocity until $t = \tau_{\mathrm{KH}}$ is reduced until a certain amount of viscosity where, instead of increasing exponentially (positive slope), the growth decreases exponentially (negative slope). We consider the amount of viscosity where the slope changes as the major indication for full suppression. For viscosities higher than this threshold the instability decays, i.e., it is fully suppressed. 

By fitting a linear function to our data we can estimate when the slope changes its sign, and therefore, the viscosity threshold\footnote{For this fit we used the time interval of [$0.05\tau_\mathrm{KH}$, $\tau_\mathrm{KH}$] to avoid the initial decay of the velocity at early times.}. We obtain that the viscosity threshold computed numerically is between $1.5\cdot10^{-3} \, \eta$ and $2\cdot10^{-3} \, \eta$. This translates to a kinematic viscosity threshold in the range of $\nu = [49.580 - 66.106]$ (see table \ref{tab:viscosity}).

For a validation of our numerical result, we estimate this viscosity threshold theoretically using three different approaches suggested by \cite{Roediger_2013}. 

The growth of the KHI described in Section~\ref{KHI} is only true for a steady background flow ($\partial v_x / \partial t = 0$), which is strictly speaking only true for low values of the viscosity. For higher values of the viscosity, we cannot assume a steady background flow anymore ($\partial v_x / \partial t \neq 0$) due to the fact that the viscosity smooths out the $x$-velocity gradient (see appendix \ref{app:viscosity}). Therefore, we cannot get an analytical solution and we need to make different assumptions to estimate this viscosity threshold. 

For the first estimate, we use the fact that the physical viscosity smooths out the velocity gradient over a length $\pm d$ above and below the interface. As demonstrated by \cite{Chandrasekhar_1961}, the KHI is suppressed for wavelengths smaller than $\sim 10d$. Now, we make use of the diffusion length $l_D = \pm 2\sqrt{\nu t}$, which measures how much the interface gets widened by diffusion at time $t$. If we take into account that in the inviscid case it takes $t = \tau_\mathrm{KH}$ for the instability to grow, we can calculate the width of the interface at that time and see whether the instability is able to grow or not. If $\lambda < 10 l_D(t=\tau_\mathrm{KH})$, the KHI will be suppressed, and otherwise it will grow. So using the definition of $\tau_\mathrm{KH}$ (equation \ref{eqn:KH_growth_time}), we obtain 
\begin{equation}
    \lambda < 10 l_D(\tau_\mathrm{KH}) = 20 \sqrt{\nu \tau_\mathrm{KH}}
\end{equation}
\begin{equation}
    \frac{\lambda^2}{400} < \nu \, \frac{\lambda}{\Delta v_x} \frac{(\rho_1 + \rho_2)}{(\rho_1\,\rho_2)^{1/2}} 
\end{equation}
\begin{equation}
    \nu > \nu_{\mathrm{Crit}} = \frac{\lambda \, \Delta v_x}{400} \frac{(\rho_1\,\rho_2)^{1/2}}{(\rho_1 + \rho_2)} \, .
\end{equation}
Using our values inferred from our initial set-up, we get a critical viscosity of $\nu_{\mathrm{Crit}} = 12.07$, which is below what we calculated numerically. However, as \cite{Roediger_2013} state, the interface is being smoothed out continuously, and therefore comparing the wavelength of the perturbation with the diffusion length at $t=\tau_\mathrm{KH}$ is somewhat arbitrary.  

For the second estimate, we assume that the effect of the viscosity dominates when the viscous dissipation time-scale, which is given by $\tau_\nu = d^2/\nu$, is shorter than the KH time-scale $\tau_\mathrm{KH}$. As mentioned before, the KHI is suppressed if $\lambda < 10d$, so we can write $d$ as $d = \lambda/10$. Now, if we compare both time-scales we get
\begin{equation}
    \tau_\mathrm{KH} > \tau_\nu
\end{equation}
\begin{equation}
    \frac{\lambda}{\Delta v_x} \frac{(\rho_1 + \rho_2)}{(\rho_1\,\rho_2)^{1/2}} > \frac{\lambda^2}{100 \nu}
\end{equation}
\begin{equation}
    \nu > \nu_{\mathrm{Crit}} = \frac{\lambda \, \Delta v_x}{100} \frac{(\rho_1\,\rho_2)^{1/2}}{(\rho_1 + \rho_2)} \, .
\end{equation}
Under these assumptions, the critical value of the viscosity is four times bigger than before, leading to $\nu_{\mathrm{Crit}} = 48.27$. This threshold correlates much better with our results and is in the range of values we measured.

Finally, we made a third estimate assuming that the instability is suppressed when it reaches its maximum height and the width of the $x$-velocity gradient is bigger than the height of the roll. The roll usually reaches a height of $\lambda/2$ at $t = \tau_\mathrm{KH}$, so this means that at $t = \tau_\mathrm{KH}$ the width of the $x$-velocity gradient must be larger than $\lambda/2$
\begin{equation}
    \frac{\lambda}{2} < l_D (\tau_\mathrm{KH}) = 2 \sqrt{\nu \tau_\mathrm{KH}}
\end{equation}
\begin{equation}
    \frac{\lambda^2}{16} < \nu \, \frac{\lambda}{\Delta v_x} \frac{(\rho_1 + \rho_2)}{(\rho_1\,\rho_2)^{1/2}}
\end{equation}
\begin{equation}
    \nu > \nu_{\mathrm{Crit}} = \frac{\lambda \, \Delta v_x}{16} \frac{(\rho_1\,\rho_2)^{1/2}}{(\rho_1 + \rho_2)} \, .
\end{equation}
This gives us a value of $\nu_{\mathrm{Crit}} = 301.70$, which is too large for our simulations. This can be due to the fact that we assumed that the maximum height is reached at $t = \tau_\mathrm{KH}$. However, by considering the top panel of Fig.~\ref{fig:amplitude_McNally_SPH}, one can see that it is reached at later times. If, for example, instead of considering that the maximum height is reached at $t = \tau_\mathrm{KH}$, we consider that it is reached at $t = 2\tau_\mathrm{KH}$, the value for the critical viscosity is reduced by half. 

Additionally, we do our estimate depending on the exact smoothing of the $x$-velocity gradient. As explained in appendix \ref{app:viscosity}, the $x$-velocity gradient is smoothed out following 
\begin{equation}
    v_x(y) = |v_{x_0}| \, \mathrm{erf} \, \left(\frac{y}{2 \sqrt{\nu t}} \right) \, .
    \label{eqn:error_function_theor}
\end{equation}
This formula arises from solving the Rayleigh problem \citep[also known as Stokes first problem;][]{Stoke_1851, Rayleigh_1911} for a viscous fluid where there are two flat plates located at the boundaries. These plates suddenly accelerate to some fixed constant velocities in opposite directions, leading to the velocity profile shown in equation \ref{eqn:error_function_theor} \citep[e.g.][]{Drazin_2006}. In our case we do not have fixed plates moving at constant velocities, which means that the particles at the boundaries will progressively slow down and the theoretical result will not be valid anymore. 

The more viscous the system is, the faster the system will move away from the initial state of the Rayleigh problem. Assuming that the instability is not suppressed if at $t = \tau_\mathrm{KH}$ the system still follows equation \ref{eqn:error_function_theor}, i.e., the particles at the boundaries still move at their initial $v_x$, one can estimate the maximum viscosity that allows this behaviour.

We also assume that our boundaries correspond to the particles at $d = \pm \lambda/2$, which is the height that the rolls are expected to reach. Only at $t = 0$ the particles move exactly at the initial $x$-velocity, so in order to do this calculation, we consider three different cases: we consider that the particles still move at their initial $v_x$ when they move $10\%$ slower than the initial $v_x$, when they move $1\%$ slower and $0.1\%$ slower. After this computation we got 
\begin{enumerate}
    \item $10\%$ slower \hspace{0.2cm}\hspace{\punt} $\rightarrow$ \hspace{0.2cm} $\nu_{\mathrm{Crit}} = 66.02$
    \item $1\%$ slower \hspace{0.2cm}\hspace{\punt}\hspace{\zero} $\rightarrow$ \hspace{0.2cm} $\nu_{\mathrm{Crit}} = 45.00$
    \item $0.1\%$ slower \hspace{0.2cm} $\rightarrow$ \hspace{0.2cm} $\nu_{\mathrm{Crit}} = 33.97$ .
\end{enumerate}

Given the values obtained by these estimates, we can see that the results we obtained for the viscosity threshold are in agreement with what we have estimated. These estimates were made using very general and ideal assumptions and we cannot rely much the exact value we got. However, we can observe that our results are in keeping with theoretical expectations.

\subsection{Energy conservation}

As the simulation runs, the friction between particles produces a loss of kinetic energy, turning it into internal energy, and since the domain is a closed system with periodic boundary conditions, the total energy has to be conserved as a function of time. To test how well the code conserves energy, we first compute the mean kinetic energy per unit mass of the whole simulation domain by summing up the kinetic energy contributed by each particles and dividing by the total number of particles. We repeat the procedure for the internal energy per unit mass of the system.

\begin{figure}
	\includegraphics[width=\columnwidth]{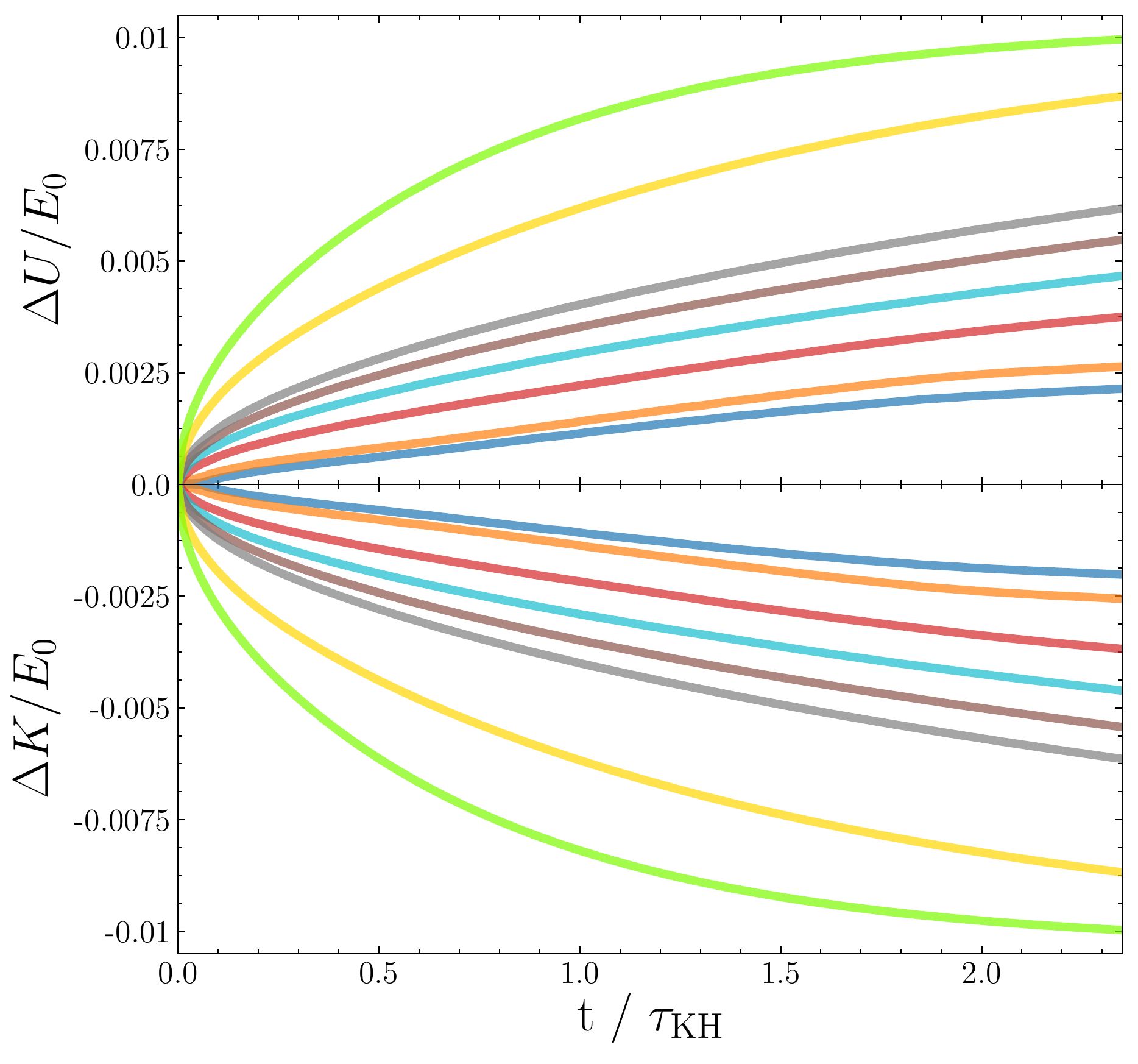}
    \caption{\textit{Top panel}: Variation of the mean internal energy per unit mass normalized to the initial total energy. \textit{Bottom panel}: Variation of the mean kinetic energy per unit mass normalized to the initial total energy. The runs with a higher viscosity transform more kinetic energy into internal energy and, the symmetry of the two plots, demonstrates the conservation of energy in all the runs. We use the same color code as the previous plots (Fig.~\ref{fig:amplitude_McNally_visc}).}
    \label{fig:Energy_visc}
\end{figure}

In Fig.~\ref{fig:Energy_visc} we show the variation of the internal (top panel) and kinetic energy (bottom panel) of the system normalized to the initial total energy for each run.
For higher values of viscosity, there is more friction between particles and more kinetic energy is turned into internal energy. The symmetry between the two panels shows the conversion of kinetic into internal energy and the conservation of energy of the system. By summing up the two values to compute the total energy we find that the runs with physical viscosity conserve the $99.992$\% of the total energy and the run without physical viscosity the $99.986$\%. We note that, while energy is largely conserved in both cases, the performance of the runs with viscosity in terms of energy conservation is slightly improved.

\subsection{Total viscosity} 

Finally, we measure the actual viscosity of the system and compare it to the Braginskii viscosity, implemented in the code. For the computation of the effective viscosity of the fluids in our simulations, we do the same analysis we made in Section~\ref{Intrinsic_visc_SPH} but with the difference that we have now a theoretical value to compare with and validate our method. After computing the total viscosity of our simulations and calculating the average value for different times, we find a small standard deviation errors, meaning that the results are consistent in time. These results can be seen in table \ref{tab:viscosity}. 

Overall, there is a good correlation between the theoretical value and the computed one. The relative errors\footnote{The relative errors are calculated dividing the absolute error by the value and multiplying by 100.} are quite low ($\sim 1\%$), with the highest error being $5.45\%$ in the less viscous case. 

\begin{figure}
	\includegraphics[width=\columnwidth]{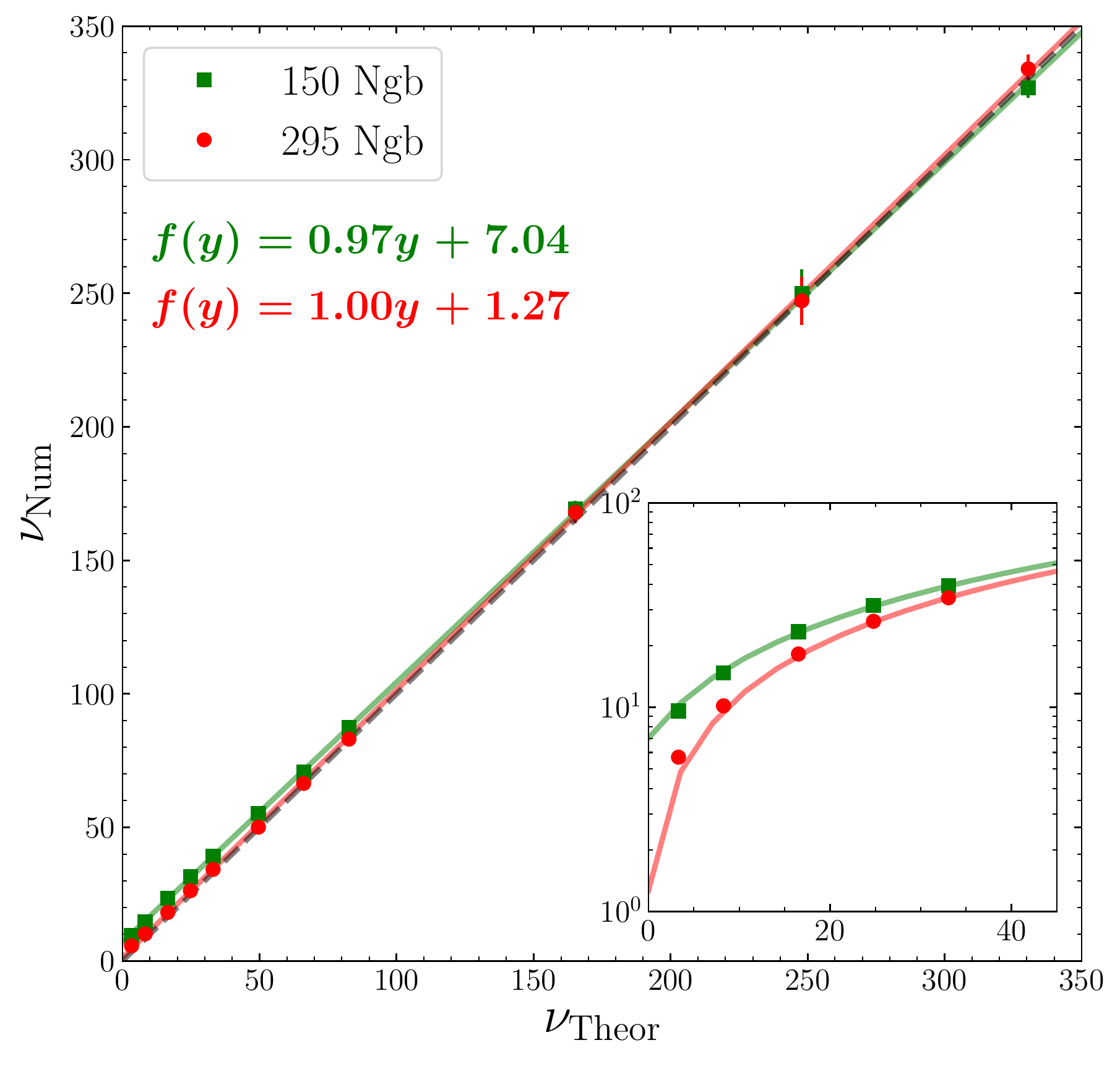}
    \caption{Numerical computation of the total viscosity of the system against the theoretical viscosity we implement. One would expect a one-to-one relation (black dashed line). In the run with $N_{\mathrm{ngb}} = 295$ the data follows a linear function with a slope of $1.001631\pm0.000013$, but an intercept of $1.3\pm0.3$, which is slightly higher than the one expected. In the case of $N_{\mathrm{ngb}} = 150$, the slope is $0.973146\pm0.000009$ and the intercept $7.04\pm0.17$, which corresponds to a shift upwards of the data.}
    \label{fig:Num_vs_theor_visc}
\end{figure}

In order to see the correlation between the theoretical (input) value and the actual (from the fit) value of the viscosity for the different runs, we plot the viscosity we obtained numerically versus the theoretical one in Fig.~\ref{fig:Num_vs_theor_visc}. For comparison, we show the cases with $N_{\mathrm{ngb}} = 150$ and $N_{\mathrm{ngb}} = 295$. Since they are expected to be the same, we would expect a relation with a slope of 1 and the intercept at $y = 0$ (black dashed line). Hence, we fit our points to a linear function. 
In the case of $N_{\mathrm{ngb}} = 295$ the data follows a linear trend with a slope of $1.001631\pm0.000013$, meaning that the growth is what we could expect, but an intercept of $1.3\pm0.3$, which means that the actual viscosity is slightly higher than the theoretical one. This discrepancy becomes more relevant in the case of $N_{\mathrm{ngb}} = 150$, where the slope is $0.973146\pm0.000009$, also close to one, but the intercept is higher than in the previous case. In this case the intercept is $7.04\pm0.17$, showing a systematic shift upwards. This systematic increase is likely due to the intrinsic numerical viscosity of the code, which is acting alongside the physical, Braginskii type viscosity that we implemented in the code. This would explain the larger viscosity for $N_{\mathrm{ngb}} = 150$ vs $N_{\mathrm{ngb}} = 295$. Note that the contribution of the intrinsic viscosity to the total viscosity of the code in the case of $N_{\mathrm{ngb}} = 150$ is similar to the value of intrinsic viscosity measured in the ideal case, but in the case of $N_{\mathrm{ngb}} = 295$ the contribution is smaller. This means that in the latter case, when the viscosity is low enough, the system will be governed by the intrinsic viscosity of the code and the physical viscosity will be negligible.

\section{Dependence on the ICs} \label{Read_IC}

In order to test how robust our results are against a change in the ICs, we adopt a new set of IC following the suggestions of \citet{Read_2010} (OG-SPH-Read and OG-MFM-Read). For completeness, we are going to compare the results obtained with both ICs to see how the set-up employed affects the growth of the KHI.

Now, the domain consists of periodic boundary conditions defined by $\Delta x = 1$, $\Delta y = 1$ and $\Delta z = 1/32$ and satisfies:
\begin{equation}
    \rho, T, v_x = \left \{ 
    \begin{matrix} \rho_1, T_1, v_1 & |y| < 0.25 \\
    \rho_2, T_2, v_2 & |y| > 0.25 \end{matrix} \right . \, .
\end{equation}
The densities and temperatures ratio $R_\rho = \rho_1 / \rho_2 = T_2 / T_1$ is equal to two, same as in the ICs given by our fiducial set-up. Since no particular density or temperature is specified, we use the same densities and temperatures we were using ($\rho_1 = 6.26\cdot10^{-8}$ and $\rho_2 = 3.13\cdot10^{-8}$; $T_1 = 2.5\cdot10^6$ and $T_2 = 5\cdot10^6$). No $x$-velocity is specified either, but the mach number is set to be $M_2 = -v_2/c_2 \approx 0.11$ and $M_1 = M_2 \sqrt{R_\rho} \approx 0.15$. Due to the fact that the mach number is given by the $x$-velocity and the speed of sound, but the speed of sound is given by the temperature, the $x$-velocities must be set to $v_1 = -26$ and $v_2 = 26$ in order to fulfil the value of the mach numbers. Despite the small length of the box in the $z$ direction, the code is written to ensure that no particle is counted twice during the neighbour finding.

The perturbation that triggers the instability is produced at the interface between the two fluids, at $y_{\mathrm{Int}} = \pm 0.25$ and is given by equation 58 in \citet{Read_2010}. The equation is equal to the one we employed to trigger the instability in previous sections (equation \ref{eqn:vy}), but the first term on the right hand side has a negative sign. This introduces a phase shift in the perturbation between the top and bottom interface. In this case, the wavelength of the perturbation is $\lambda = 0.5$, the scale parameter $\sigma$ remains the same, $\sigma = 0.2\lambda$ and the initial amplitude of the perturbation is $\delta v_y = |v_x|/8 = 3.25$. A summary of all the differences can be seen in table \ref{tab:Read_IC}.
\begin{table*}
    \centering
    \caption{Differences between the ICs used for triggering the KHI in OG-SPH and OG-MFM \citep{Murante_2011} and the ones in OG-SPH-Read and OG-MFM-Read \citep{Read_2010}.}
    \renewcommand\tabcolsep{5.mm}
    \begin{tabular}{|c|c|c|c|c|}
        \hline
         & Box Size & $v_x$ & $v_y$ (eq. \ref{eqn:vy}) & Mach Number \\
        \hline \hline
        $\begin{matrix}
            \text{OG-SPH} \\ \text{OG-MFM}
        \end{matrix} 
        $ & $256\times256\times8$ & $
            \begin{matrix}
            v_{x_1} = -40 \\ v_{x_2} = +40
            \end{matrix}
        $ & $
            \begin{matrix}
            \lambda = 128 \\ \sigma = 0.2\lambda \\ \delta v_y = |v_x|/10 = 4 \\ y_{\mathrm{Int}} = \pm 64
            \end{matrix}
        $ & $
            \begin{matrix}
            M_1 \approx 0.23 \\ M_2 \approx 0.17
            \end{matrix}
        $ \\
        \hline
        $\begin{matrix}
            \text{OG-SPH-Read} \\ \text{OG-MFM-Read}
        \end{matrix} 
        $ & $1\times1\times1/32$ & $
            \begin{matrix}
            v_{x_1} = -26 \\ v_{x_2} = +26
            \end{matrix}
        $ & $
            \begin{matrix}
            \lambda = 0.5 \\ \sigma = 0.2\lambda \\ \delta v_y = |v_x|/8 = 3.25 \\ y_{\mathrm{Int}} = \pm 0.25
            \end{matrix}
        $ & $
            \begin{matrix}
            M_1 \approx 0.15 \\ M_2 \approx 0.11
            \end{matrix}
        $ \\
        \hline
    \end{tabular}
    \label{tab:Read_IC}
\end{table*}

\begin{figure*}
    \centering
	\includegraphics[width=\textwidth]{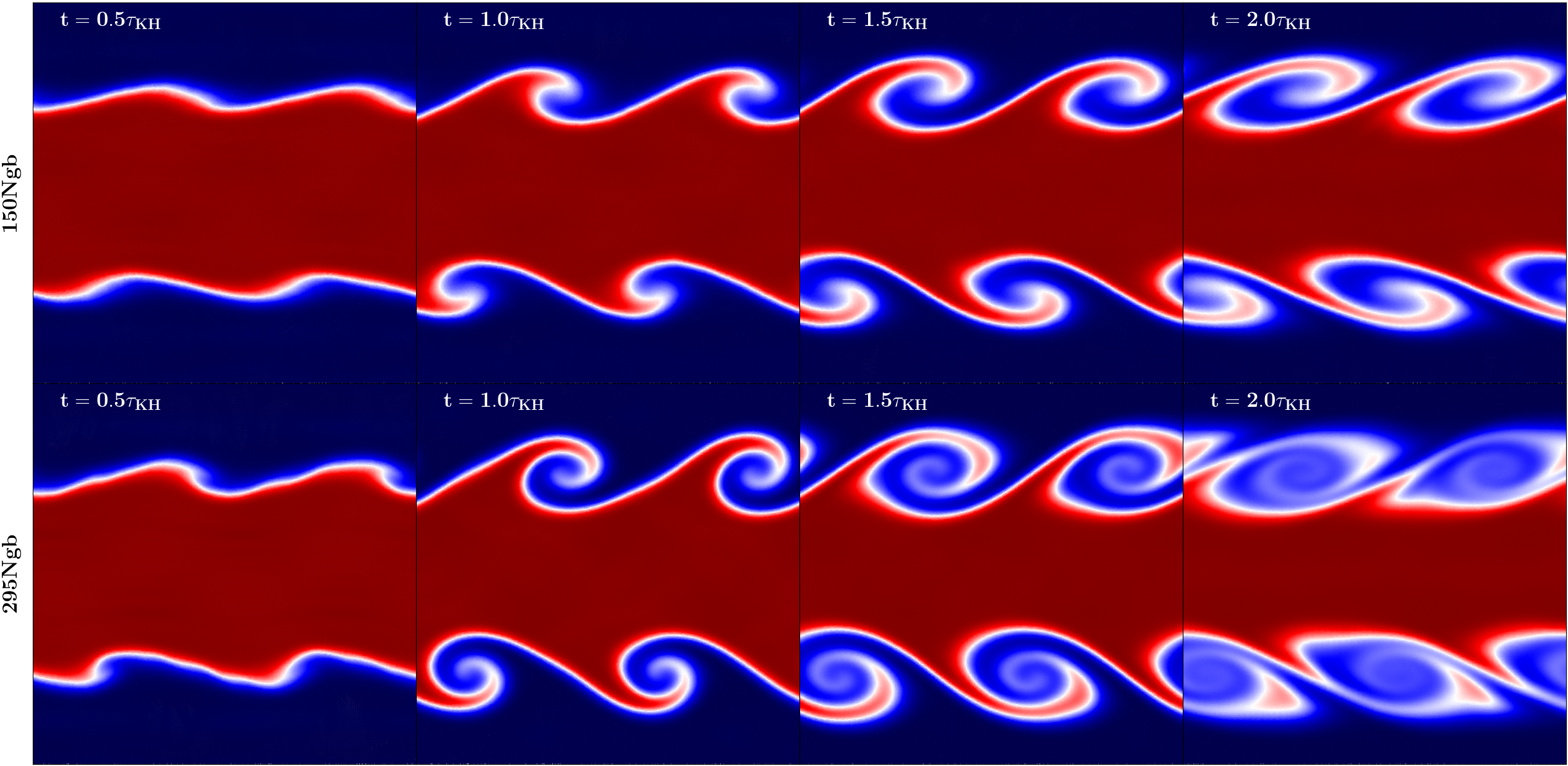}
    \caption{Colormap of the densities using OG-SPH-Read. In this case the behaviour with $N_{\mathrm{ngb}} = 295$ is similar to the ones with OG-SPH (see Fig.~\ref{fig:colormaps_SPH}), but using $N_{\mathrm{ngb}} = 150$ the instability cannot develop as much as with OG-SPH.}
    \label{fig:colormaps_Read}
\end{figure*}

For a qualitative comparison of the results, Fig.~\ref{fig:colormaps_Read} shows the column density of the runs with OG-SPH-Read and $N_{\mathrm{ngb}} = 150$ and $N_{\mathrm{ngb}} = 295$ (top and bottom row respectively). The shape of the rolls with $N_{\mathrm{ngb}} = 295$ does not differ much from the ones obtained with OG-SPH (see Fig.~\ref{fig:colormaps_SPH}), while we can see a difference in the results using $N_{\mathrm{ngb}} = 150$. In contrast to OG-SPH, the instability does not achieve the roll shape obtained with OG-SPH. It is also worth noting the shift between the top and the bottom rolls due to the phase shift of the $y$-velocity on the top and bottom interface in the ICs.

Due to the small distance between the interfaces, the flow might not remain constant away from the instability and, therefore, the top instabilities could in principle affect the bottom ones and vice versa. However, in appendix \ref{app:large} we show that top and bottom instabilities do not affect each other growth.

\subsection{Results for SPH}

\begin{figure}
    \centering
	\includegraphics[width=\columnwidth]{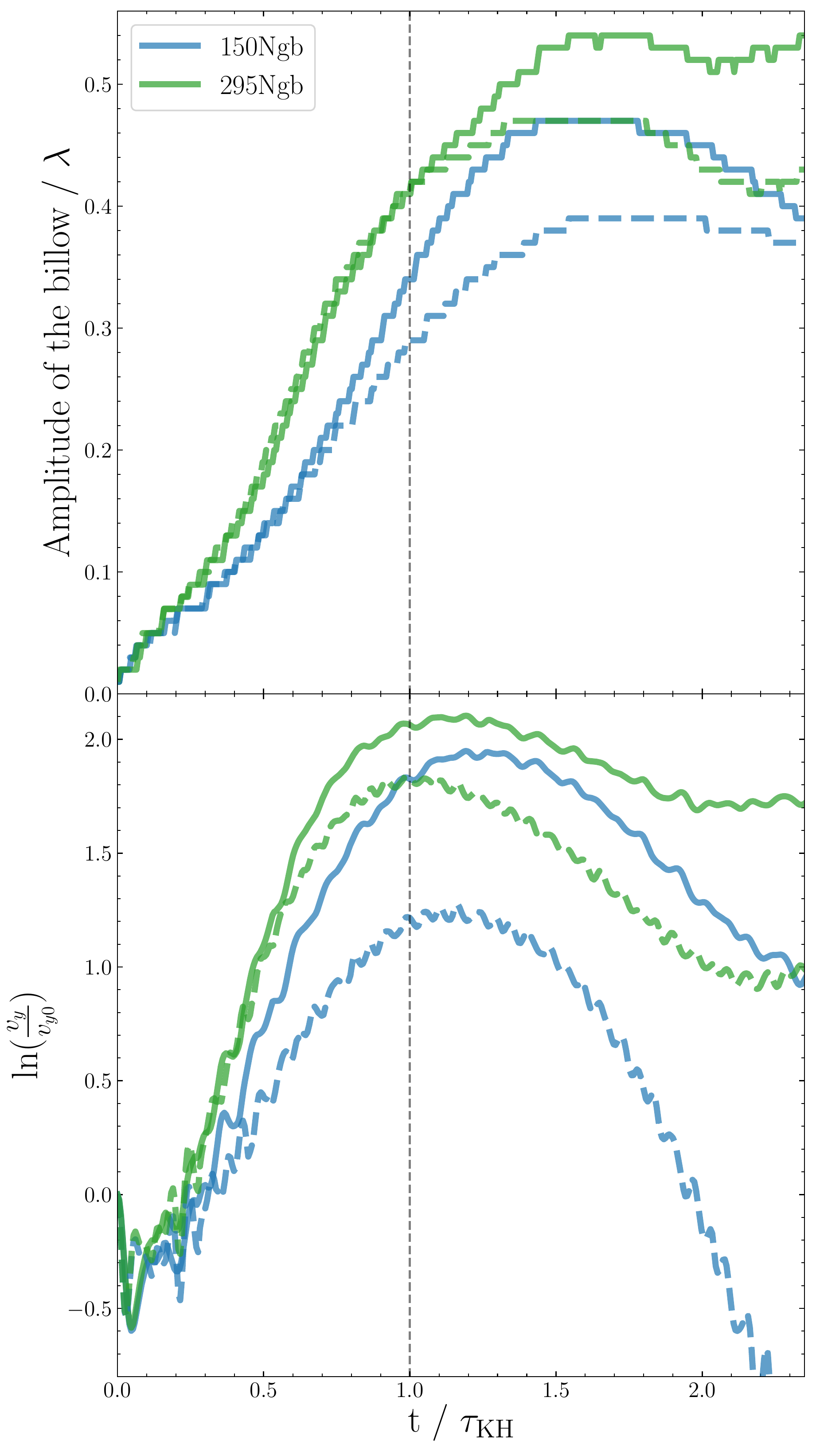}
    \caption{\textit{Top panel}: Height of the rolls with $N_{\mathrm{ngb}} = 150$ and $N_{\mathrm{ngb}} = 295$ using OG-SPH (solid lines) and OG-SPH-Read (dashed lines). The maximum amplitude reached with $N_{\mathrm{ngb}} = 295$ is close to $\lambda/2$ in both cases, but the one reached with $N_{\mathrm{ngb}} = 150$ is smaller with OG-SPH-Read than with OG-SPH. \textit{Bottom panel}: Evolution of the $y$-velocity for the cases with $N_{\mathrm{ngb}} = 150$ and $N_{\mathrm{ngb}} = 295$ with OG-SPH (solid lines) and OG-SPH-Read (dashed lines). The performance of $N_{\mathrm{ngb}} = 150$ with the new ICs is worse than with our ICs, showing a slower growth of the instability. The results with $N_{\mathrm{ngb}} = 295$ show a faster growth using OG-SPH. Although the slower growth of OG-SPH-Read, the rolls reach a maximum amplitude close to $\lambda/2$.}
    \label{fig:SPH_Read}
\end{figure}

To obtain a more quantitative comparison between the two different set-ups, we plot in the top panel of Fig.~\ref{fig:SPH_Read} the change of the height of the rolls with time for both cases: solid lines represent the runs with OG-SPH and the dashed lines the ones with OG-SPH-Read. Both set-ups show a similar growth of the amplitude depending on $N_{\mathrm{ngb}}$ at early times but, in agreement with the qualitative analysis done before, the maximum amplitude reached by OG-SPH-Read with $N_{\mathrm{ngb}} = 150$ is smaller (below $0.4 \lambda$) compared to the runs using OG-SPH (close to $\lambda/2$). The run employing OG-SPH-Read and $N_{\mathrm{ngb}} = 295$ reaches a maximum amplitude similar to the run with OG-SPH and $N_{\mathrm{ngb}} = 150$, which is close to the expected value of $\lambda/2$. Despite both runs with $N_{\mathrm{ngb}} = 295$ follow a similar path until $t = \tau_{\mathrm{KH}}$, the case using OG-SPH is able to reach a higher maximum amplitude compared to the run using OG-SPH-Read.

The way in which the height of the rolls evolves with time can be explained with the results for the growth of the $y$-velocity, which we show in the bottom panel of Fig.~\ref{fig:SPH_Read}. For the case of $N_{\mathrm{ngb}} = 150$ the instability grows faster and for a longer time with OG-SPH than with OG-SPH-Read, which explains why it reaches a higher amplitude. In the case of $N_{\mathrm{ngb}} = 295$, the rate of growth is very similar at early times, which explains that the growth of the amplitude in both cases is very similar until $t = \tau_{KH}$. Then the run with OG-SPH-Read stabilises, while the one with OG-SPH keeps growing, reaching a larger velocity. The growth of the $y$-velocity with OG-SPH and $N_{\mathrm{ngb}} = 150$ is less steep but grows for a longer time compared to OG-SPH-Read with $N_{\mathrm{ngb}} = 295$, which explains why the amplitude grows slower at early times but reaches the same maximum height.

\subsection{Results for MFM}

\begin{figure}
	\includegraphics[width=\columnwidth]{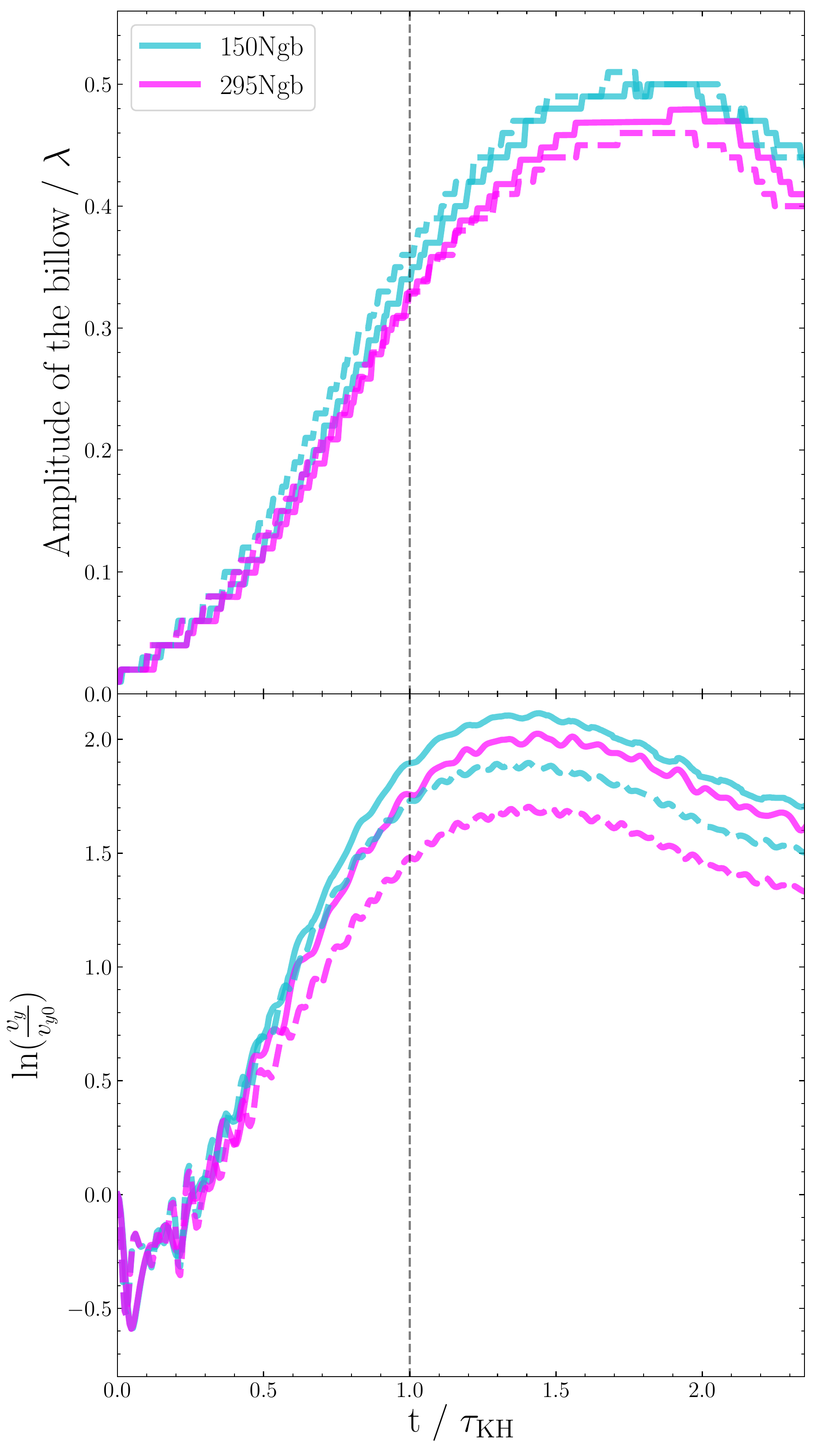}
    \caption{\textit{Top panel}: Growth of the amplitude of the rolls in the case of OG-MFM and OG-MFM-Read with $N_{\mathrm{ngb}} = 150$ and $N_{\mathrm{ngb}} = 295$. In both cases the behaviour is very similar independently of the initial set-up we use to trigger the KHI. \textit{Bottom panel}: Evolution of the $y$-velocity with time for the runs using OG-MFM and OG-MFM-Read with $N_{\mathrm{ngb}} = 150$ and $N_{\mathrm{ngb}} = 295$. The growth of the $y$-velocity differs at later times between the two set-ups, where the runs using OG-MFM reach higher velocities than the cases with OG-MFM-Read. Additionally, the differences in the results depending on $N_{\mathrm{ngb}}$ are larger with the OG-MFM-Read set-up than with OG-MFM.}
    \label{fig:MFM_Read}
\end{figure}

Finally, we show the results for OG-MFM in the top panel of Fig.~\ref{fig:MFM_Read} for the evolution of the height of the billows for the simulations made with $N_{\mathrm{ngb}} = 150$ and $N_{\mathrm{ngb}} = 295$. In this case, in contrast to SPH, the results for the amplitude with both set-ups follow a very similar behaviour, where the discrepancy is negligible. In both cases the run with $N_{\mathrm{ngb}} = 150$ evolves faster and reaches a larger amplitude than the run with $N_{\mathrm{ngb}} = 295$. Despite the behaviour of the height of the rolls is similar with both set-ups, the bottom panel of Fig.~\ref{fig:MFM_Read} shows that the growth of the $y$-velocity is similar at early times, but then it differs at later times between the two different ICs. The runs with OG-MFM reach higher velocities than OG-MFM-Read for both $N_{\mathrm{ngb}} = 150$ and $N_{\mathrm{ngb}} = 295$, although this does not trigger a substantial difference in the height of the rolls between the two set-ups. Comparing the results depending on $N_{\mathrm{ngb}}$ in each set-up, we find that the differences in OG-MFM-Read for the two cases are larger than the ones obtained in OG-MFM.

Overall, we find that OG-SPH with a low $N_{\mathrm{ngb}}$ ($N_{\mathrm{ngb}} = 150$) can be sensitive to the ICs and lead to different results depending on the set-up employed, while with a higher $N_{\mathrm{ngb}}$ ($N_{\mathrm{ngb}} = 295$) the results appear to be similar independently of the set-up. In the case of OG-MFM, the results of the amplitude of the rolls are very similar independently of the ICs and $N_{\mathrm{ngb}}$, but the evolution of the $y$-velocity shows larger discrepancies between the two set-ups.

\begin{figure}
    \centering
	\includegraphics[width=\columnwidth]{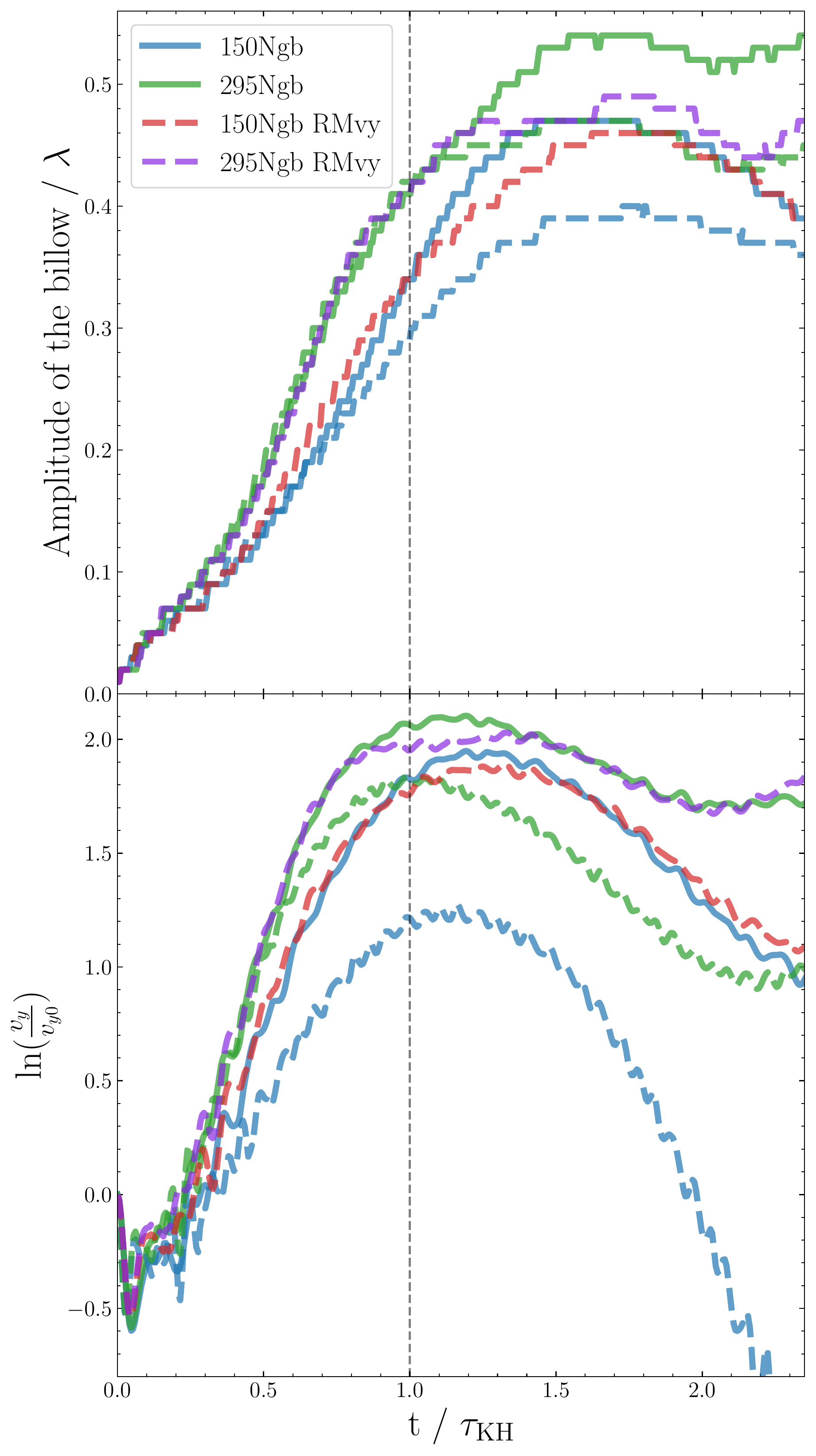}
    \caption{\textit{Top panel}: Growth of amplitude with $N_{\mathrm{ngb}} = 150$ and $N_{\mathrm{ngb}} = 295$ with OG-SPH set-up (solid lines), the original OG-SPH-Read set-up (blue and green dashed lines) and the RMvy set-up (red and violet dashed lines). \textit{Bottom panel}: Evolution of the $y$-velocity with OG-SPH set-up (solid lines), the original OG-SPH-Read set-up (blue and green dashed lines) and the RMvy set-up (red and violet dashed lines). When using RMvy, we recover the original behaviour of OG-SPH in both the amplitude evolution and the growth of the $y$-velocity.}
    \label{fig:Amplitude_McNally_both}
\end{figure}

To study the origin of these differences between OG-SPH-Read and OG-MFM-Read, we run the simulations again but using a different set-up. We employed the ICs suggested in \citet[][]{Read_2010}, but we used $v_{x_1} = \pm -40, v_{x_2} = 40$ and $\delta v_y = |v_x|/10 = 4$ (labeled as RMvy). This means that we used the same Mach number and the same amplitude of the perturbation employed in OG-SPH and OG-MFM. After running the simulations, we recover the behaviour found with OG-SPH in the linear regime (see Fig. \ref{fig:Amplitude_McNally_both}). This suggests that the velocity of the fluids can modify the intrinsic viscosity of the code in our SPH scheme, while no significant change is found when using MFM. This makes us think that our MFM implementation is more stable against thermal to kinetic energy ratio variations than SPH.

\section{Conclusions} \label{Summary}

In this work we carried out a detailed investigation of fluid mixing comparing our implementation of SPH and MFM in the code \textsc{OpenGadget3}. First, we tested the ability of the code to capture the KHI. Then, we performed different simulations after including physical viscosity to the system to analyse the behaviour of the KHI in viscous fluids. Additionally, we changed the original set-up and studied the impact of the ICs on final results for both SPH and MFM. Our key conclusions are:
\begin{itemize}
\item In all the runs with OG-SPH the instability can fully develop the characteristic roll of the KHI, which is visible in all cases. However, the growth rate when the highest $N_{\mathrm{ngb}}$ is assumed is faster than in the case with $N_{\mathrm{ngb}} = 150$ and $N_{\mathrm{ngb}} = 200$. In the latter two cases, a lower maximum amplitude is reached, which is however still close to the maximum expected value \citep[$\sim \lambda/2$,][]{Roediger_2013}. In the cases with $N_{\mathrm{ngb}} = 295$ and $N_{\mathrm{ngb}} = 350$ the billows manage to grow higher than $\lambda/2$, and we can observe the fastest growth rate for the KHI.

By computing the diffusion of the code we could observe that a constant AC in OG-SPH might be too diffusive, while a TDAC reduces the diffusion to the minimum needed to trigger the KHI. In order to understand how viscous are the systems that we simulated, we measured the intrinsic viscosity of the code. We showed that, for the case with $N_{\mathrm{ngb}} = 150$, the fluids are more viscous compared to the other runs, which could explain why the KHI grows slower; while with $N_{\mathrm{ngb}} \geq 250$ the intrinsic viscosity reaches a minimum and remains stable for higher $N_{\mathrm{ngb}}$. Despite the higher viscosity found in the case with $N_{\mathrm{ngb}} = 150$, the overall viscosity allows in all cases the complete development of the instability. We showed that this intrinsic viscosity has nothing to do with the artificial viscosity implemented in OG-SPH, which is successfully suppressed at early times independently of $N_{\mathrm{ngb}}$.

\item In agreement with OG-SPH, the results obtained with OG-MFM fulfil the expectations independently of $N_{\mathrm{ngb}}$. However, one needs at least $N_{\mathrm{ngb}} = 150$ in order to successfully suppress the growth of the secondary instabilities. Above $N_{\mathrm{ngb}} = 150$ the evolution of the amplitude of the rolls and the $y$-velocity is very similar for all the cases studied with OG-MFM. The growth of the KHI is slower with OG-MFM than with OG-SPH: as a consequence, the maximum amplitude is reached at later times. This is due to the fact that the intrinsic viscosity in OG-MFM is higher than in the cases with OG-SPH. However, this excess of intrinsic viscosity does not suppress the instability. By analysing how diffusive OG-MFM is, we found that it is less diffusive than OG-SPH with a constant AC, but the behaviour is similar to the run with OG-SPH and TDAC. In conclusion, one needs at least $N_{\mathrm{ngb}} = 150$ to successfully reproduce the KHI but it must be taken into account that OG-MFM is computationally more expensive than OG-SPH and, the higher $N_{\mathrm{ngb}}$, the more expensive it becomes.

\item To test the Braginskii viscosity implemented in \textsc{OpenGadget3} we computed again the amplitude of the billows using OG-SPH, showing that the instability cannot grow in a highly viscous fluid, but the height of the rolls increases when decreasing the amount of viscosity. By measuring the evolution of the $y$-velocity we found a viscosity threshold ($\approx10^{-3} \, \eta$) where, for viscosities higher than this threshold, the KHI is fully suppressed and for smaller amounts of viscosity the instability is able to grow exponentially. We computed the threshold numerically and compared it with three different theoretical estimates and one numerical estimate, concluding that the threshold computed numerically is in agreement with these estimates. In terms of energy conservation, the higher the viscosity in the simulations, the more kinetic energy is turned into internal energy. In this process the code is conserving always more than $99.99$\% of the total energy. Additionally, we measured the actual viscosity of the system and compared it to the theoretical viscosity we had implemented, observing that the effective viscosity of the system tends to be higher than the one we implement. This effect could be explained if the intrinsic viscosity of the code is taken into account, meaning that the total viscosity of the system we simulate is not only the physical viscosity we implement, and the intrinsic viscosity of the code contributes as well. This contribution is also dependent on $N_{\mathrm{ngb}}$, where the case with $N_{\mathrm{ngb}} = 150$ shows a larger contribution compared to the run with $N_{\mathrm{ngb}} = 295$.

\item The run with OG-SPH and $N_{\mathrm{ngb}} = 150$ is very sensitive to the initial set-up employed to trigger the KHI. In the case with OG-SPH-Read the rolls cannot grow as much as they do with OG-SPH and the roll shape does not fully develop. Despite the rolls obtained with OG-SPH and $N_{\mathrm{ngb}} = 295$ reach a higher amplitude, the ones with OG-SPH-Read manage to reach a height close to $\lambda/2$. With OG-MFM-Read, however, the behaviour is similar independently of the set-up employed to trigger the instability.

\item A change in the initial velocity of the fluids introduces a modification in the intrinsic viscosity of the code, which happens to be more sensitive in OG-SPH than in OG-MFM. This means that OG-MFM is more stable against variations of the thermal to kinetic energy ratio than OG-SPH.
\end{itemize}

In summary, \textsc{OpenGadget3} successfully reproduces the linear growth of the KHI using different hydro solvers with different numerical and physical set-ups. We find that the changes of the inferred numerical viscosity in our different set-ups of our SPH implementation are comparable to the differences between the SPH and MFM results. In general the SPH results are more sensitive to the details of the set-up and it is recommended to use time dependent artificial conduction (TDAC) in order to avoid over-mixing. Nevertheless, SPH reproduces the expected reduction of the growth rate in the presence of physical viscosity and recovers very well the threshold level of physical viscosity needed to fully suppress the instability. In the case of galaxy clusters with a virial temperature of $3\times10^7$K, this level corresponds to a suppression factor of $\approx10^{-3}$ of the classical Braginskii value. The intrinsic, numerical viscosity of our SPH implementation is found to be only half the value obtained for the MFM implementation; within an ICM environment, this corresponds to a value smaller by an order of magnitude (i.e. $\approx10^{-4} \, \eta$). All the tests presented are re-ensuring that modern SPH methods are suitable to study the effect of physical viscosity in galaxy clusters.

\section*{Acknowledgements}
The authors thank Frederick Groth for in advance access to the MFM solver in \textsc{OpenGadget3}. The authors also want to thank the referee for their very useful comments. TM, MV and KD are supported by the Excellence Cluster ORIGINS which is funded by the Deutsche Forschungsgemeinschaft (DFG, German Research Foundation) under Germany´s Excellence Strategy – EXC-2094 – 390783311.
MV acknowledges support from the Alexander von Humboldt Stiftung and the Carl Friedrich von Siemens Stiftung. UPS is supported by the Simons Foundation through a Flatiron Research Fellowship (FRF) at the Center for Computational Astrophysics. The Flatiron Institute is supported by the Simons Foundation. KD acknowledges funding for the COMPLEX project from the European Research Council (ERC) under the European Union’s Horizon 2020 research and innovation program grant agreement ERC-2019-AdG 882679. 
UPS would like to thank Eve C. Ostriker for the intense discussion about the KHI in particle codes which inspired this project in the first place. 
TM, MV, UPS and KD acknowledge the computing time provided by the Leibniz Rechenzentrum (LRZ) of the Bayrische Akademie der Wissenschaften on the machine SuperMUC-NG (pr86re). UPS acknowledges the computing time provided by the Leibniz Rechenzentrum (LRZ) of the Bayrische Akademie der Wissenschaften on the machine SuperMUC-NG (pn72bu). We thank the super computing resources at the LRZ in Garching for using an energy mix that is to $100$ per cent comprised out of renewable energy resources (e.g. \url{https://www.top500.org/news/germanys-most-powerful-supercomputer-comes-online/}, \\\url{https://www.lrz.de/wir/green-it_en/}).

\section*{Data Availability}
The data underlying this article will be shared on reasonable request to the author.
 



\bibliographystyle{mnras}
\bibliography{Bibliography}




\appendix

\section{Amplitude of the rolls} \label{app:amplitude}

The method employed to compute the height of the rolls is similar to the one described in \citet{Roediger_2013}. We focus only on the upper half of the domain (since the domain is symmetric, we choose only the upper half, and therefore, the total amount of particles is reduced by half). Then we mark every particle depending if they are ``red'' (denser fluid) or ``blue'' (lighter fluid) at $t=0$ so we can trace them later. Once we have marked every particle, we divide the half domain in 100 bins along the $y$ direction and calculate the amount of ``red'' and ``blue'' particles in every bin for every snapshot. The top of the billow will correspond to the lowest bin where at least 95\% of the particles are ``blue'', while the bottom will be the highest bin with at least 95\% of ``red'' particles. Finally, we compute the amplitude of the roll by calculating the distance between the top and the bottom of the billow for every snapshot.

\section{Growth of the velocity} \label{app:v_y}

We use the discrete convolution suggested by \cite{McNally_2012} to compute the amplitude $M$ of the initially excited mode. We have adapted the formula to our ICs, leading to
\begin{equation}
    s_i = v_y h_i^3 \, \sin \left(\frac{2\pi \left(x+\frac{\lambda}{2}\right)}{\lambda}\right) \, \exp \left(-\frac{2\pi}{\lambda} \, |64 - y| \right)
\end{equation}
\begin{equation}
    c_i = v_y h_i^3 \, \cos \left(\frac{2\pi \left(x+\frac{\lambda}{2}\right)}{\lambda}\right) \, \exp \left(-\frac{2\pi}{\lambda} \, |64 - y| \right)
\end{equation}
\begin{equation}
    d_i = h_i^3 \, \exp \left(-\frac{2\pi}{\lambda} \, |64 - y| \right)
\end{equation}
\begin{equation}
    M = 2 \, \sqrt{\left(\frac{\sum_{i=1}^N s_i}{\sum_{i=1}^N d_i} \right)^2 + \left(\frac{\sum_{i=1}^N c_i}{\sum_{i=1}^N d_i} \right)^2} \, ,
\end{equation}
where $h_i$ is the smoothing length, $\lambda$ the wavelength of the perturbation and $N$ is the total number of particles in the domain. For the computation of $M$ we used only one quarter of the full domain in order to take only one perturbation for the calculation.

\section{Computation of diffusion} \label{app:diffusion}

\begin{figure}
    \centering
    \includegraphics[width=\columnwidth]{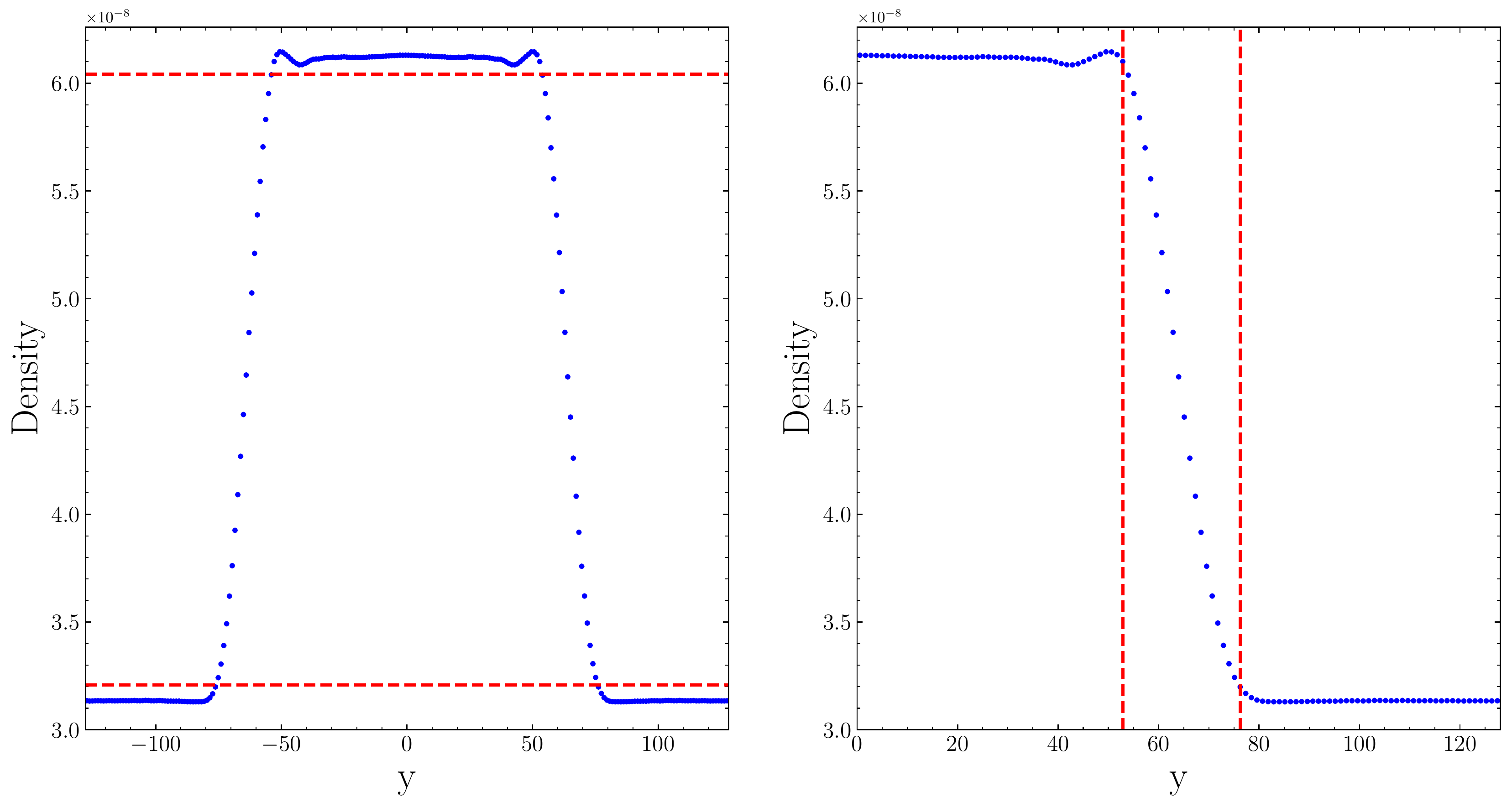}
    \caption{\textit{Left panel}: Plot of density against $y$ position at $t = 0.4 \tau_\mathrm{KH}$, where the upper red dashed line indicates the minimum density for a particle to be considered `dense' and the lower red dashed line the maximum density for a particle to be considered `light'. \textit{Right panel}: Plot of the density against $y$ position for positive values of $y$ also at $t = 0.4 \tau_\mathrm{KH}$, where the two vertical red dashed lines indicate the width of the interface.} 
    \label{fig:profile_diff}
\end{figure}

We set up the same ICs like in our KHI box, but this time without ad-hoc seeded perturbation. Then we divide the whole initial density domain in 20 bins and we choose the bins with the highest and lowest density. These are going to be our thresholds to consider if a particle belongs to the high density part or to the low density one (see left plot of Fig.~\ref{fig:profile_diff}). Now, for each snapshot, we take the positive values of $y$, divide them in 115 bins and compute the mean density in each bin. Then we consider the width of the interface to be the distance between the rightmost `dense' bin and the leftmost `light' bin (see right plot of Fig.~\ref{fig:profile_diff}). At later times some numerical instabilities can grow, and therefore, affect our results, so we computed the diffusion until $t \sim 0.4 \tau_\mathrm{KH}$.

\section{Intrinsic viscosity of the system} \label{app:viscosity}

The effect of viscosity is to smooth out the velocity gradient between the two fluids by momentum diffusion, and therefore, the more viscosity a system has, the more the gradient is smoothed out and the more difficult it is to develop the instability. The $x$-velocity profile is smoothed out following
\begin{equation}
    v_x(y) = |v_{x_0}| \, \mathrm{erf} \, \left(\frac{y}{2 \sqrt{\nu t}} \right) \, ,
    \label{eqn:error_function}
\end{equation}
where the interface is set at $y=0$, $|v_{x_0}|$ is the initial $x$-velocity of one of the fluids (the two fluids have the same speed but in opposite directions) and $\nu$ is the kinematic viscosity of the system. In order to calculate the intrinsic viscosity of the system, we simulate our two fluids without any initial perturbation and we fit the analytical function \ref{eqn:error_function} to our data at different times with the kinematic viscosity as a free parameter. To do so, we use only the top half of the full domain and we displace it to set the interface at $y=0$. We then divide the half domain in 127 bins and compute the mean $x$-velocity of each bin at five different times. Finally, we calculate the average value of the five fits in order to get a value for the total viscosity of the system.
\begin{figure}
	\includegraphics[width=\columnwidth]{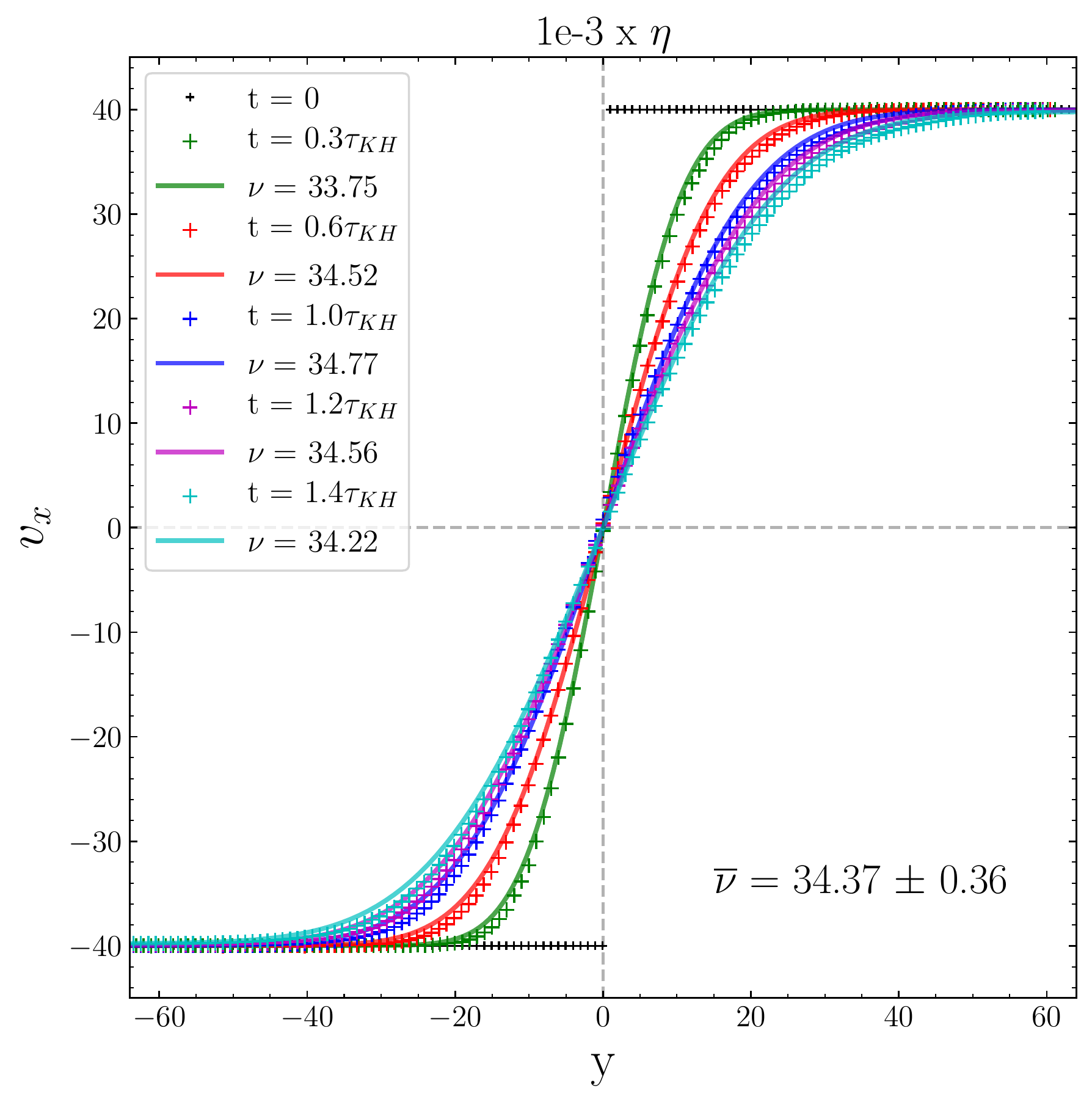}
    \caption{Fit of the analytic formula \ref{eqn:error_function} to our data for the computation of the kinematic viscosity for the different simulations. It starts from a discontinuity in the $x$-velocity profile at $t = 0$ and, as time passes, the $x$-velocity gradient is smoothed out by viscosity. The crosses represent our data and the solid lines the fit of the analytic function, coded by different colors for different times.}
    \label{fig:fit_visc}
\end{figure}

\section{Values of intrinsic viscosity for OG-SPH and OG-MFM} \label{app:table_vis}

The values obtained after computing the intrinsic viscosity depending on $N_{\mathrm{ngb}}$ for OG-SPH and OG-MFM are shown in table \ref{tab:intrinsic_visc}. These values are plotted in figure \ref{fig:Intrinsic_visc}. 
\begin{table}
    \centering
    \caption{Values of the intrinsic viscosity of the codes depending on $N_{\mathrm{ngb}}$.}
    \renewcommand\tabcolsep{7.mm}
    \begin{tabular}{|c||c|c|}
        \hline
         & \multicolumn{2}{c}{Intrinsic Viscosity} \\
        \hline
        $N_{\mathrm{ngb}}$ & OG-SPH & OG-MFM \\
        \hline \hline
        150 & $6.78\pm0.38$ & $8.62\pm1.47$ \\
        \hline
        200 & $4.73\pm0.57$ & $9.49\pm1.32$ \\
        \hline
        250 & $4.10\pm0.83$ & $11.50\pm2.00$ \\
        \hline
        295 & $3.78\pm0.73$ & $10.50\pm2.07$ \\
        \hline
        350 & $4.30\pm0.96$ & $11.74\pm2.07$ \\
        \hline
    \end{tabular}
    \label{tab:intrinsic_visc}
\end{table}

\section{Cubic spline kernel in OG-MFM} \label{cubic_spline}

\begin{figure}
    \centering
	\includegraphics[width=\columnwidth]{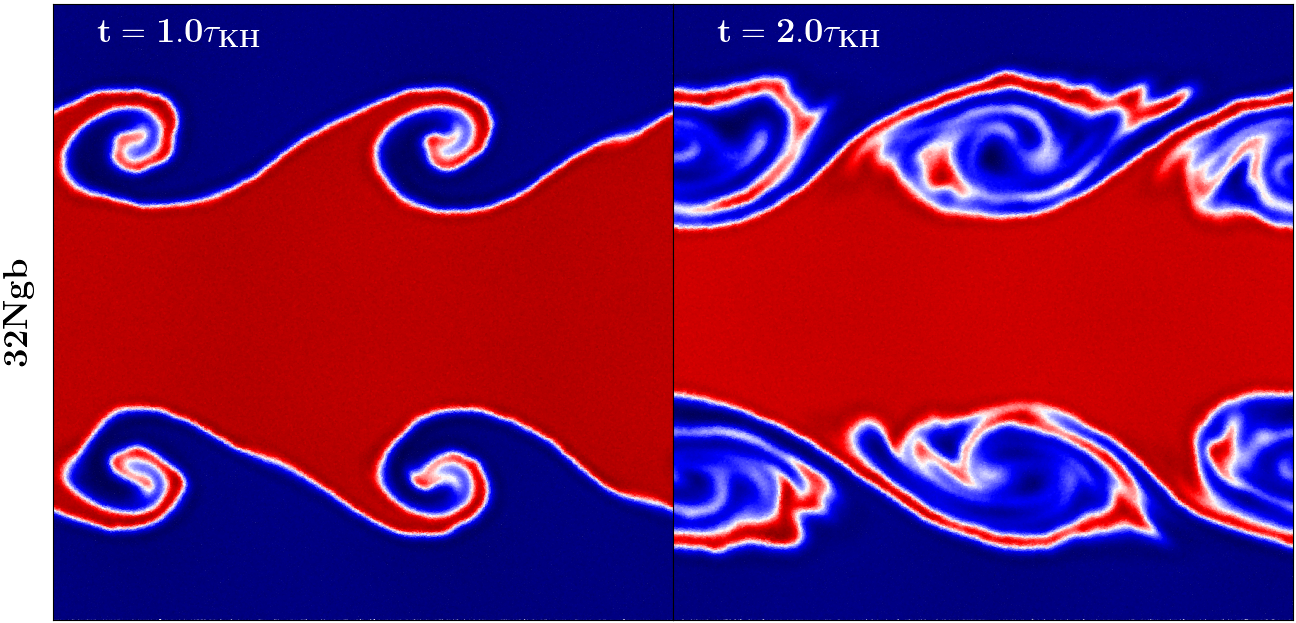}
    \caption{Colormaps of the density from the simulation run with OG-MFM using a cubic spline kernel and $N_{\mathrm{ngb}} = 32$. Results shown for $t = \tau_\mathrm{KH}$ and $t = 2 \tau_\mathrm{KH}$.}
    \label{fig:colormaps_MFM_cubic_spline}
\end{figure}

The `E$_0$ error' does not take place in MFM simulations, so in order to check the behaviour of the KHI with a lower $N_{\mathrm{ngb}}$, we additionally run a simulation using OG-MFM with a cubic spline kernel and $N_{\mathrm{ngb}} = 32$. As Fig.~\ref{fig:colormaps_MFM_cubic_spline} shows, the KHI can grow successfully and the mixing of the fluids takes place as expected. However, although it is computationally less expensive, the secondary instabilities are not properly suppressed and are able to grow leading to a non-fully symmetric result, pushing away the simulations from the expected result \citep[see][for reference]{Robertson_2010, McNally_2012}.

\section{Double-size box using OG-SPH-Read} \label{app:large}

\begin{figure}
    \centering
	\includegraphics[width=\columnwidth]{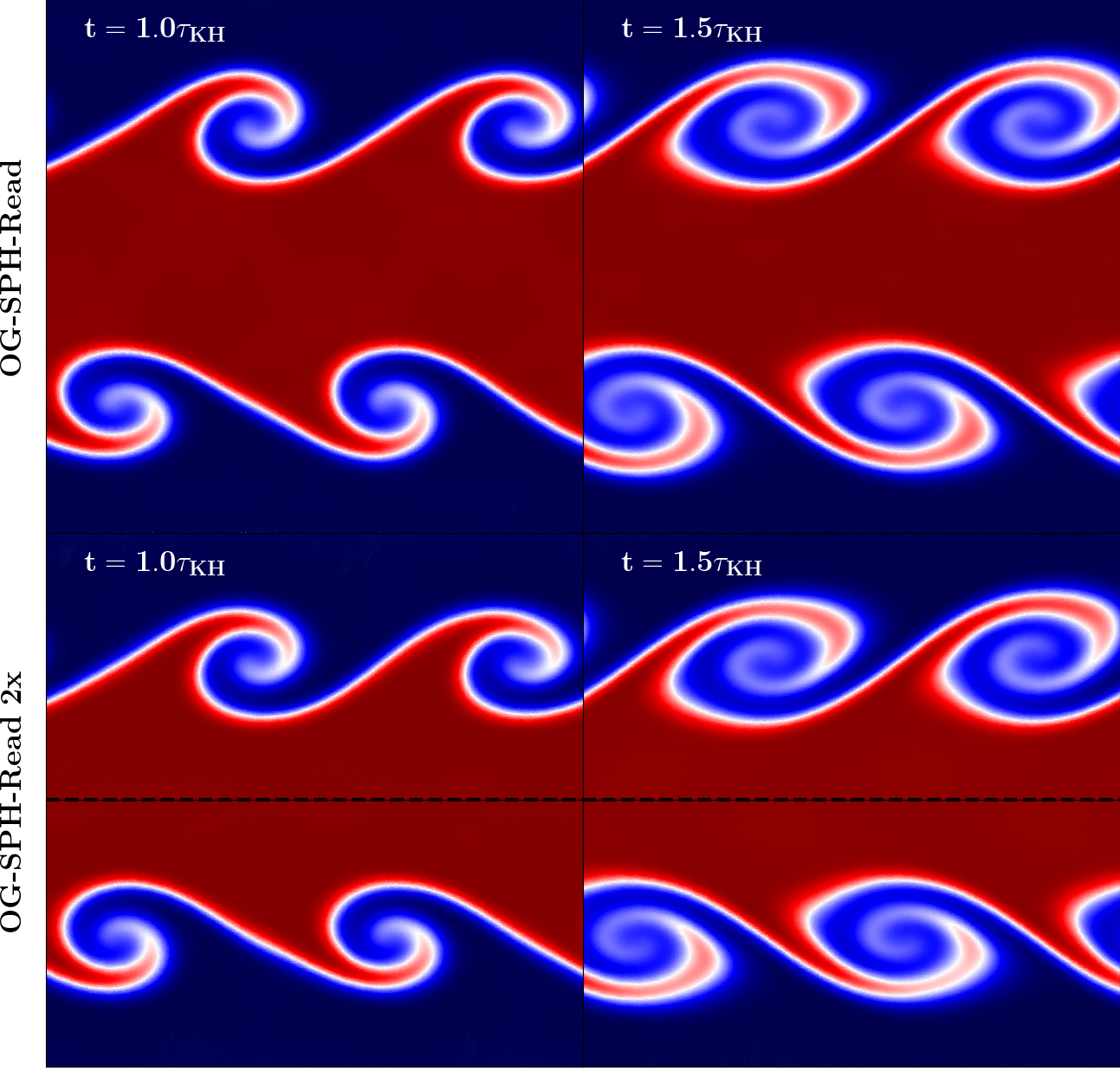}
    \caption{Colormaps of the density from the simulation run with OG-SPH-Read (top row) and using OG-SPH-Read 2x (bottom row), which corresponds to a box two times larger in the $y$ direction. Both simulations using $N_{\mathrm{ngb}} = 295$. Despite the different sizes of the boxes, the same region has been plotted for comparison. The black dashed line in the bottom row indicates where the cut-off has been made.}
    \label{fig:colormaps_large}
\end{figure}

\begin{figure}
    \centering
	\includegraphics[width=\columnwidth]{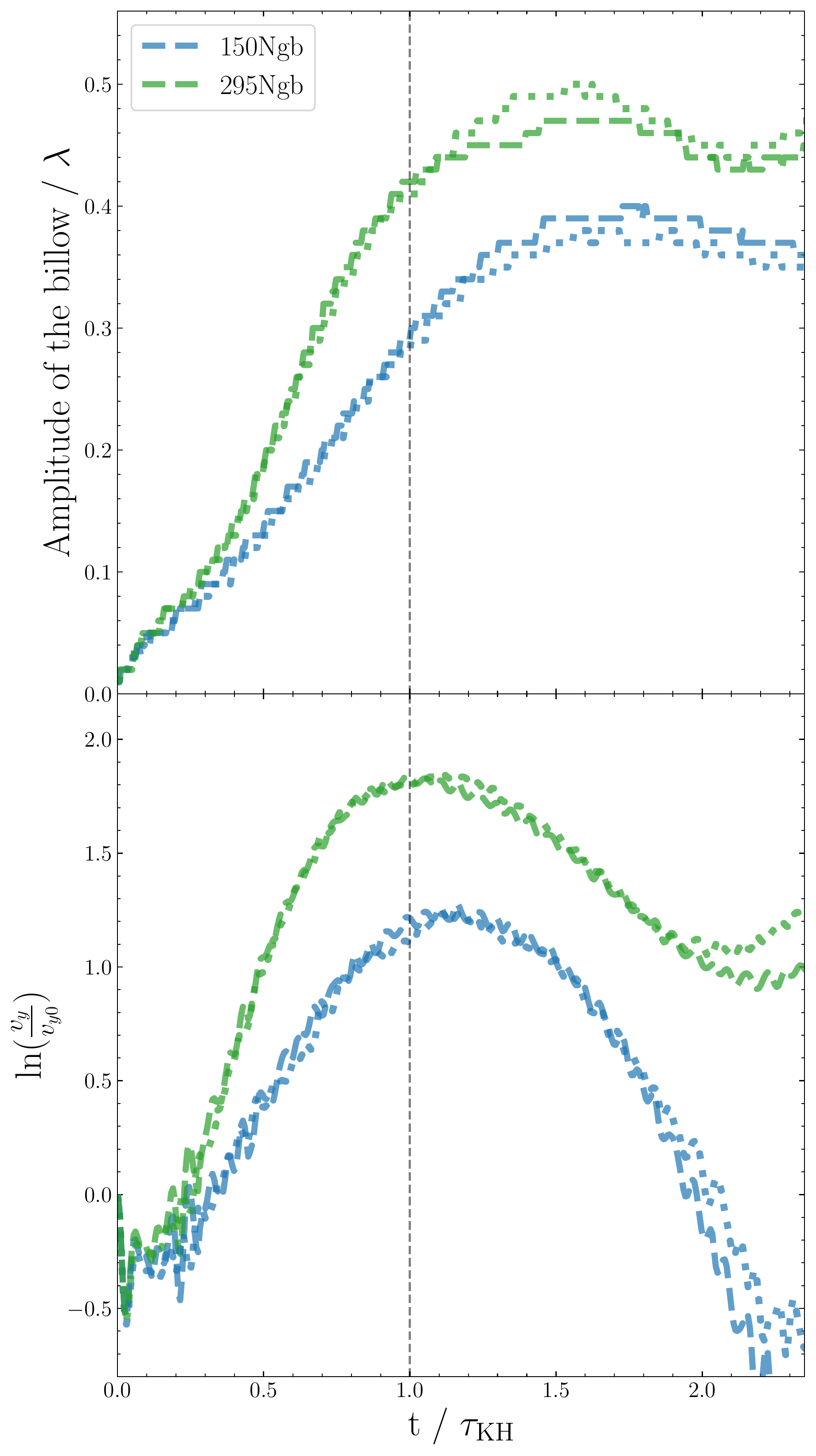}
    \caption{Growth of amplitude and velocity with time with $N_{\mathrm{ngb}} = 150$ and $N_{\mathrm{ngb}} = 295$ with the original OG-SPH-Read set-up (dashed lines) and with a box two times larger in the $y$ direction (dotted lines).}
    \label{fig:Amplitude_McNally_2x}
\end{figure}

To test whether the top instabilities affect the bottom instabilities and vice versa, we rerun the simulations with OG-SPH-Read, but doubling the distance between the two contact discontinuities. The resulting box has a length in the $y$ axis of $\Delta y = 2$ and the interfaces are set at $y = \pm 0.5$. The colormap in Fig.~\ref{fig:colormaps_large} shows the original run with the OG-SPH-Read set-up using $N_{\mathrm{ngb}} = 295$ (top row) and its counterpart with the larger box (bottom row, labeled as OG-SPH-Read 2x). The same regions as the original set-up has been plotted for comparison (the cut off has been marked with a black dashed line). The growth of both instabilities shows a very similar shape, suggesting that the instabilities of the top and bottom interfaces are not affecting each other. This is supported by figure \ref{fig:Amplitude_McNally_2x}, where we plot a comparison of the growth of the amplitude and velocity (dashed lines OG-SPH-Read and dotted lines OG-SPH-Read 2x). The growth of the amplitude and $v_y$ follows the same path in both set-ups for the linear regime (which is what we are interested in in this paper). In the non-linear regime some differences can be seen, but overall the behaviour is very similar. The same results can be observed using OG-SPH.

\bsp	
\label{lastpage}
\end{document}